\documentclass[twocolumn,english,aps,prd,reprint,floatfix,notitlepage,footinbib,preprintnumbers,superscriptaddress,altaffilletter]{revtex4-2}

\usepackage{amsmath,amssymb,amsfonts}
\usepackage{hyperref,breakurl,cleveref,url}
\usepackage{color}

\usepackage{graphicx}
\usepackage[export]{adjustbox}

\usepackage{accents,mathrsfs,mathtools}

\renewcommand{\paragraph}[1]{%
    \textit{#1}.---%
}

\def\skip{\vskip1.5pt}
\newcommand\trick[1]{}

\usepackage{enumitem}
\setlist[enumerate]{
    label={},
    leftmargin=2em,
    itemsep=2pt,
    topsep= 2pt,
    partopsep=0pt,
    parsep=0pt,
}

\let\oldeqref\eqref
\renewcommand{\eqref}[1]{Eq.\,\smash{\oldeqref{#1}}}
\newcommand{\eqrefs}[2]{Eqs.\,\smash{\oldeqref{#1}}\,and\,\smash{\oldeqref{#2}}}

\newcommand{\rcite}[1]{Ref.\,\cite{#1}}
\newcommand{\rrcite}[1]{Refs.\,\cite{#1}}

\newcommand{\fref}[1]{Fig.\,\ref{#1}}

\def\mem{\hspace{0.1em}}
\def\hem{\hspace{0.05em}}
\def\nem{\hspace{-0.1em}}
\def\hnem{\hspace{-0.05em}}
\def\hhem{\hspace{0.025em}}
\def\hhnem{\hspace{-0.025em}}
\def\hhhem{\hspace{0.0125em}}

\def\blank{{\,\,\,\,\,}}

\def\qiq{{\quad\implies\quad}}

\def\a{\alpha}
\def\b{\beta}
\def\c{{\gamma}}

\def\m{\mu}
\def\n{\nu}
\def\r{\rho}
\def\s{\sigma}
\def\k{\kappa}
\def\l{\lambda}
\def\t{\tau}

\def\mdot{{\mem\cdot\mem}}

\newcommand{\wrap}[1]{{\smash{#1}\vphantom{\beta}}}

\def\lsq{{
    \kern-0.037em
    \adjustbox{scale=0.919,valign=c}{$
        {
            \adjustbox{raise=-0.0855em}{$\lfloor$}
            \llap{\reflectbox{\rotatebox[origin=c]{180}{$\lfloor$}}}
        }
    $}
    \kern-0.04em
}}
\def\rsq{{
    \kern-0.04em
    \adjustbox{scale=0.919,valign=c}{$
        {
            \rlap{\reflectbox{\rotatebox[origin=c]{180}{$\rfloor$}}} 
            \adjustbox{raise=-0.0855em}{$\rfloor$}
        }
    $}
    \kern-0.037em
}}

\def\mflat{\mathbb{M}}

\def\M{{\mathcal{M}}}
\def\P{{\mathcal{P}}}
\def\mathe{{\scalebox{1.05}[1.01]{$\mathrm{e}$}}}
\def\sprime{{\mathrlap{\smash{{}^\prime}}{\hspace{0.05em}}}}

\def\i{{\iota}}

\newcommand{\co}[1]{c^{\hhem#1}}
\newcommand{\cop}[1]{ \mathrlap{c'}\phantom{c}^{\kern0.24em#1} }

\usepackage{tikz}
\usetikzlibrary{calc} % to use relative coordinates
\usetikzlibrary{shapes.geometric} % to draw regular polygons
\usetikzlibrary{positioning} % to use right=of 
\usetikzlibrary{fit} % for fit size
\usepackage[a]{esvect} % arrow styling %f
\tikzset{empty/.style = {inner sep = 0pt, outer sep = 0, minimum size = 0}}
\tikzset{b/.style = {inner sep = 2pt, outer sep = 4pt, minimum size = 12pt}}
\tikzset{w/.style = {inner sep = 1pt, outer sep = 2pt, minimum size = 12pt, anchor = west}}
\tikzset{s/.style = {inner sep = 2.5pt, outer sep =2.5pt, minimum size = 1pt, font = \small}}

\definecolor{sky}{RGB}{144,187,231}
\definecolor{OxyRed}{RGB}{190,70,62}
\definecolor{NitroBlue}{RGB}{91,122,239}
\definecolor{HydrogenLight}{RGB}{245,250,252}

\tikzset{
	line/.style = {draw, line width = 1.1pt, line cap = round, rounded corners = 0.2pt},
	bine/.style = {draw, line width = 1.1pt, line cap = round, rounded corners = 0.0pt, dotted, color=OxyRed},
	dine/.style = {draw, line width = 1.4pt, line cap = round, rounded corners = 0.0pt, double},
	D/.style = {below = -1.1pt, font = \footnotesize\bfseries\sffamily},
	U/.style = {above = -1.2pt, font = \footnotesize\bfseries\sffamily},
	L/.style = {left,  font=\footnotesize},
	R/.style = {right, font=\footnotesize},
	X/.style = {circle, draw=black, fill=HydrogenLight, inner sep=0pt, outer sep=0pt, minimum size=4.5pt, line width=1.1pt},
	Y/.style = {circle, draw=black, fill=NitroBlue, inner sep=0pt, outer sep=0pt, minimum size=4.5pt, line width=1.1pt}
}

\usepackage[export]{adjustbox}

\newcommand{\bb}[1]{\bigg(\,{#1}\,\bigg)}
\newcommand{\BB}[1]{\Big(\,{#1}\,\Big)}
\newcommand{\bigbig}[1]{\big(\mem{#1}\mem\big)}

\newcommand{\bbsq}[1]{\bigg[\,{#1}\,\bigg]}

\newcommand{\lrp}[1]{\left(\mem{#1}\mem\right)}
\newcommand{\lrsq}[1]{\left[\mem{#1}\mem\right]}

\def\mdot{{\mem\cdot\mem}}

\def\swedge{{\mem{\wedge}\,}}

\def\mplus{{\mem+\mem}}

\def\mminus{{\mem-\mem}}

\def\L{\mathcal{L}}

\def\O{\mathcal{O}}

\def\acX{\acute{X}}
\def\acY{\acute{Y}}
\def\acQ{\acute{Q}}

\def\id{{\rlap{1} \hskip 1.6pt \adjustbox{scale=1.1}{1}}}

\def\D{\text{\fontfamily{lmtt}\fontseries{b}\selectfont D}}

\newcommand{\hls}[1]{{\color[RGB]{0,137,186}{}#1}}

\begin{document}

\title{
	Geodesic Deviation to All Orders
	via a Tangent Bundle Formalism
}

\author{Joon-Hwi Kim}
\affiliation{Walter Burke Institute for Theoretical Physics, California Institute of Technology, Pasadena, CA 91125}

\begin{abstract}
    We establish an in-in formalism for geodesic deviation
    as an alternative to Synge calculus,
	based on a covariant calculus of differential forms in tangent bundle.
    This derives
    the exact Lagrangian and equations
    governing the finite geodesic deviation between
    a free-falling test particle
    and an arbitrary observer,
    in terms of
    infinite sums
    whose coefficients are products of binomial coefficients.
	Explicit expressions are provided up to tenth order,
	finding agreements with the previous fourth-order result.
    % 
%	Comments are given for the relevance to
%	the all-orders-in-spin dynamics of Kerr black hole
%	via 
%	a probe counterpart of
%	Newman-Janis algorithm.
\end{abstract}

\preprint{CALT-TH 2025-004}

% \date{\today}

\bibliographystyle{utphys-modified}

\renewcommand*{\bibfont}{\fontsize{8}{8.5}\selectfont}
\setlength{\bibsep}{1pt}

\maketitle

\paragraph{Introduction}% 
The geodesic deviation equation (GDE)
is a foundational topic in general relativity,
commonly covered in textbooks
\cite{Carroll:2004st,mtw,Hawking:1973uf,Wald:1984rg,PoissonAndWill2014,synge1952tensor,synge1960general}.
First derived in 1927
by Levi-Civita and Synge
\cite{levicivita1927ecart,synge1927ii,synge1928geodesics,Synge:1935zz,synge1934deviation},
this equation describes the evolution of the \textit{infinitesimal} separation between two nearby free-falling test particles.
Explicitly, it reads
\begin{align}
    \label{eq:GDE2}
    \frac{D^2 y^\m}{d\t^2}
    \,=\,
        - R^\m{}_{\n\r\s}\mem u^\n y^\r u^\s
    \,,
\end{align}
where $y^\m$ describes the infinitesimal separation as a vector
and $u^\m$ describes the unit-normalized four-velocity.

The generalization of the GDE to the case of \textit{finite} separations
has been a topic of research in the literature
\cite{hodgkinson1972modified,bazanski1977kinematics,bazanski1977dynamics,aleksandrov1979geodesic,li1979coupled,mashhoon1975tidal,mashhoon1977tidal,ciufolini1986generalized,ciufolini1986measure,Chicone:2002kb,Chicone:2006rm,Perlick:2007ux,Vines:2014oba,Puetzfeld:2015uxi,Obukhov:2018nnt,Flanagan:2018yzh,Waldstein:2021olw}.
Here, the objective is to find corrections to the GDE
perturbatively in the orders of the separation vector $y$,
covariantly defined as a tangent to the geodesic segment
joining the two particles.
Namely,
the right-hand side of \eqref{eq:GDE2}
is augmented
with terms involving
nonlinear powers of $y$,
coupled to
the Riemann tensor $R^\m{}_{\n\r\s}$ and its derivatives.
Physically, this captures
the higher-order tidal effects
due to the non-infinitesimal size of $y$.

The higher-order extensions of the GDE
have found fruitful physical applications
\cite{Vines:2014oba,Waldstein:2021olw}:
gravitational wave detectors,
astrophysical jets,
measurement of spacetime curvature,
and
an analytical treatment of
eccentric relativistic orbits
\cite{li1979coupled,mashhoon1975tidal,mashhoon1977tidal,ciufolini1986generalized,ciufolini1986measure,Chicone:2002kb,Chicone:2006rm,Perlick:2007ux,%
Kerner:2001cw,vanHolten:2001ea,Colistete:2002ka,colistete2002higher,Baskaran:2003bx,Koekoek:2010pv,Koekoek:2011mm,tammelo1984physical,tammelo2006pressure}.
For instance,
\rcite{Baskaran:2003bx} remarks that
employing a higher-order version of the GDE
could have improved the accuracy of LIGO's original data analysis by 10\% 
\cite{Vines:2014oba}.

The explicit GDE valid up to
$\mathcal{O}(y^3)$ and $\mathcal{O}(y^4)$
were obtained by
works \cite{hodgkinson1972modified,bazanski1977kinematics,bazanski1977dynamics,aleksandrov1979geodesic}
in the 1970s
and Vines \cite{Vines:2014oba}
in 2014,
respectively.
Moreover,
Vines \cite{Vines:2014oba}
also provided
formulae for
the all-orders extension of the GDE
and its Lagrangian formulation,
though
in terms of
tensor expressions
that are not fully expanded out in terms of
Riemann curvature and its derivatives.
To find the explicit equation or Lagrangian at an order,
one has to solve
a group of interrelated recursion relations
in Appendix 3 of \rcite{Vines:2014oba},
which arise
in the context of
various and intricate identities of
the Synge bitensor formalism
\cite{Ruse:1931ht,Synge:1931zz,Synge:1960ueh,Poisson:2011nh}.

In this paper,
we intend to revisit 
this problem
from an different framework.
The achievements are the following.
Firstly,
we develop an alternative 
to the Synge formalism 
in which
the relevant tensor expressions
are
computed 
from
a covariant calculus of differential forms.
Secondly, we provide
the exact all-orders formula
for the so-called Jacobi propagators
\cite{Dixon:1970zza,Dixon:1974xoz,Harte:2008xt,Harte:2012uw},
fully expanded out in terms of Riemann tensor and its derivatives.
Our exact expressions are infinite sums
whose coefficients are products of binomial coefficients,
showing
agreements with Vines \cite{Vines:2014oba}.
Consequently,
we provide
the explicit GDE and its Lagrangian up to $\O(y^{10})$
in the ancillary file \texttt{All.nb}
and Appendix~\ref{APP-L2}.

The key idea of our approach
is to formulate 
geodesic deviation 
as an initial-value problem
like in
works
\cite{Thiemann:1995ug,Thiemann:2000bw,Thiemann:2002vj,Hall:2001jq,Hall:2002du,hall2011adapted}.
Namely, we specify an \textit{in-in} boundary condition for the geodesic:
point $x$ and a tangent vector $y$ at $x$.
This is to be contrasted with
the Synge bitensor formalism
\cite{Ruse:1931ht,Synge:1931zz,Synge:1960ueh,Poisson:2011nh}
where a geodesic segment is characterized by
its two endpoints
$x$ and $z$
as an \textit{in-out} boundary condition.

\skip
\paragraph{Geodesic Deviation in Tangent Bundle}% 
Let $(\M,g)$ be
a $d$-dimensional real-analytic manifold
with local coordinates $x^\m$
and metric $g_{\m\n}$,
which we call spacetime.
Its tangent bundle, $T\M$,
can be viewed as a $2d$-dimensional manifold
with local coordinates $(x^\m,y^\m)$,
based on
its local trivialization
by the coordinate vector fields.

Coordinate transformations of $T\M$
are restricted in the form
$(x^\m,y^\m) \mapsto (f^\m(x),f^\m{}_{,\n}(x)\mem y^\n)$,
where $f^\m(x)$ describes a set of real-analytic functions in a local patch of $\M$.
Importantly,
the notion of
covariance in $T\M$ is defined with respect to
these coordinate transformations.

In particular,
the following vector field 
in $T\M$
is \textit{invariant} under 
such coordinate transformations:
\begin{align}
    \label{eq:N}
    N
    \,=\,
        y^\m \frac{\partial}{\partial x^\m}
        - \Gamma^\m{}_{\r\s}(x)\mem y^\r\mem y^\s\mem \frac{\partial}{\partial y^\m}
    \,,
\end{align}
where $\Gamma^\m{}_{\r\s}(x)$ denote the Christoffel symbols.
This can be easily seen by considering its interior products with
a one-form basis in $T\M$ that transform covariantly:
\begin{align}
    \label{eq:iN}
    \i_N\hem dx^\m = y^\m
    \,,\quad
    \i_N Dy^\m = 0
    \,.
\end{align}
Here, $D$ denotes the covariant exterior derivative:
$Dy^\m = dy^\m + \Gamma^\m{}_{\n\r}(x)\mem y^\n\mem dx^\r$.
In light of its invariance,
$N$ defines a structure
characteristic of the tangent bundle.
In the mathematical jargon,
it describes a horizontal vector field 
due to the Ehresmann notion of a connection
\cite{ehresmann1948connexions,Mason:2013sva}.

\begin{figure}[t]
    \centering
    \includegraphics[scale=1.25]{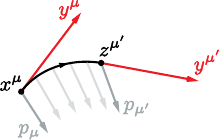}
    \caption{
        A geodesic segment joins two spacetime points $x$ (observer) and $z$ (test particle).
        The tangent vectors are respectively denoted as $y^\m$ and $y^{\m\sprime}$, 
        which are normalized such that
        $x$ and $z$ are related by unit-time geodesic flows.
    }
    \label{fig:GDprime}
\end{figure}

Crucially,
$N$ 
holds the very significance as
the \textit{generator of geodesic deviation}.
To see this, consider
the first-order formulation of the geodesic equation:
\begin{align}
\begin{split}
    \label{eq:diffeq-XY}
    \frac{d}{d\eta}\mem X^\m(\eta) 
    &\mem=\mem 
        Y^\m(\eta)
    \,,\\
    \frac{d}{d\eta}\mem Y^\m(\eta)
    &\mem=\mem
        -\Gamma^\m{}_{\r\s}(X(\eta))\,
        Y^\r(\eta)\, Y^\s(\eta)
    \,.
\end{split}
\end{align}
Evidently,
the explicit components of $N$ in \eqref{eq:N}
encode \eqref{eq:diffeq-XY}.
This implies that
the power series solution of \eqref{eq:diffeq-XY}
is given by
$X^\m(\eta) = \mathe^{\eta N} x^\m$
and
$Y^\m(\eta) = \mathe^{\eta N} y^\m$
for initial conditions
$X^\m(0) = x^\m$ and $Y^\m(0) = y^\m$,
where
$N$ is understood as a differential operator.
Namely,
the time-$\eta$ flow of $N$ solves the geodesic equation.

In particular, consider the unit-time flow:
\begin{align}
    \label{eq:XY(1)}
    X^\m(1) \mem=\mem \mathe^{N} x^\m
    \,,\quad
    Y^\m(1) \mem=\mem \mathe^{N} y^\m
    \,.
\end{align}
Borrowing the notation in the Synge bitensor formalism
\cite{Ruse:1931ht,Synge:1931zz,Synge:1960ueh,Poisson:2011nh},
we rephrase \eqref{eq:XY(1)} as the following:
\begin{align}
    \label{eq:zy}
    z^{\m\sprime}
    \mem=\mem
        \delta^{\m\sprime}{}_\m\mem
            (\mathe^{N} x^\m)
    \,,\quad
    y^{\m\sprime}
    \mem=\mem
        \delta^{\m\sprime}{}_\m\mem
            (\mathe^{N} y^\m)
    \,,
\end{align}
where $\delta^{\m\sprime}{}_\m$ is Kronecker delta.
Namely, we assign unprimed and primed indices
to objects at the original ($x$) and deviated ($z$) points,
respectively.
The rationale might be that indices represent the behavior under coordinate transformations,
and 
tensors at $x$ and $z$ transform with different Jacobian factors.
\fref{fig:GDprime} provides a spacetime picture
that visualizes \eqref{eq:zy}.

The unit-time flow of $N$
can be regarded as a diffeomorphism in $T\M$.
By the very definition of the Lie derivative,
the pullback of 
a tensor
$\a$ in $T\M$
is given by $\mathe^{\pounds_N} \a$.
Practically, this pullback
replaces
every $x$ with $\mathe^N x$ and $y$ with $\mathe^N y$.
For instance, the following identity holds:
\begin{align}
    \label{eq:ELN.dx}
    dz^{\m\sprime}
    &\mem=\mem
        \delta^{\m\sprime}{}_\m\mem
            (\mathe^{\pounds_N} dx^\m)
    \,.
\end{align}
Notably,
for differential forms,
Lie derivatives are neatly computed
by using the Cartan magic formula:
\begin{align}
    \label{eq:magic}
    \pounds_N
    \mem=\mem
        d\mem \i_N
        + \i_N d
    \,.
\end{align}
We leave it 
as an exercise to
check \eqref{eq:ELN.dx}
at each order in $y$
by using \eqref{eq:magic}.

\skip
\paragraph{Geodesic Deviation in A Direct Sum Bundle}% 
For pedagogical reasons, let us also consider
a slight generalization of the above construction.
Suppose the direct sum of
the tangent and cotangent bundles:
$\P = T\M \oplus T^*\nem\M$,
where local coordinates are $(x^\m,y^\m,p_\m)$.
Eventually, this $3d$-dimensional space
will serve as our geometrical arena
for a first-order formulation of GDE.
In this larger bundle,
the vector field $N$ is uniquely defined by
$\i_N\hem dx^\m = y^\m$,
$\i_N Dy^\m = 0$,
and
$\i_N Dp_\m = 0$.

It is not difficult to see that this $N$ is the generator of geodesic deviation and parallel transport.
In particular,
consider the analog of the first-order differential equations in \eqref{eq:diffeq-XY}
to find that
$p_{\m\sprime} = (\mathe^{N} p_\m)\mem \delta^\m{}_{\m'}$
is the covector $p_\m$
parallel-transported to the point $z$
along the geodesic;
see \fref{fig:GDprime}.
This finding is succinctly stated as
\begin{align}
    \label{eq:W.p}
    p_{\m\sprime}
    \,=\, 
        p_\m\mem W^\m{}_{\m'}
    \,,
\end{align}
where $W^\m{}_{\m\sprime}$ denotes the \textit{parallel propagator}:
the Wilson line of the Levi-Civita connection
computed about the geodesic between $x$ and $z$.
Namely,
$p_\m \mapsto p_{\m\sprime} = p_\m\mem W^\m{}_{\m'}$
describes the isomorphism between the cotangent spaces
at $x$ and $z$
facilitated by the geodesic parallel transport.

Again,
$(x,y,p) \mapsto \mathe^{\eta N}(x,y,p)$
describes the time-$\eta$ flow in $\P$ generated by the vector field $N$.
Due to
this geometrical interpretation,
the unit-time geodesic deviation and parallel transport 
is given by the
exponentiated Lie derivative $\mathe^{\pounds_N}$.

For a concrete example,
consider a one-form $p_\m\hem dx^\m$,
i.e., the trivial extension of the cotangent bundle's tautological one-form
to $\P$.
Then we have
\begin{align}
    \label{eq:pdz0}
    p_{\m'}\hem dz^{\m\sprime}
    \mem=\mem
        \mathe^{\pounds_N} (p_\m\mem dx^\m)
    \,,
\end{align}
the right-hand side of
which can be computed order-by-order
by series-expanding $\mathe^{\pounds_N}$.
We leave it as an exercise to carry out this computation
and check consistency between the left-hand and right-hand sides.

\skip
\paragraph{Covariant Lie Derivative and Dressing Identity}% 
The computations utilizing $\pounds_N$,
however,
are not very efficient.
One encounters unoccupied connection coefficients
arising from the components of $N$,
so covariance at the original point $x$
is not manifested
in intermediate steps.

In this light, we introduce a shorthand notation,
\begin{align}
    \label{eq:cov-magic}
    \pounds_N^D
    \mem=\mem
        D\mem \i_N
        + \i_N D
    \,,
\end{align}
defined on any tensor-valued differential form.
This describes a covariantized analog of the Lie derivative;
more information might be found in
\rrcite{Jackiw:2001jb,Ashtekar:2020xll,de2015covariant}.

Crucially,
since $p_\m\hem dx^\m$ 
is a scalar-valued one-form carrying no free indices,
using $D$ instead of $d$ makes no difference.
Hence we can equivalently use $\exp({\pounds_N^D})$
in \eqref{eq:pdz0}.
It is a privilege of exponentiated operators
that they are distributable as
\begin{align}
    \label{eq:pdz-dist-cov}
    \mathe^{\pounds_N^D}(p_\m\hem dx^\m)
    \hem=\hem
    (\mathe^{\pounds_N^D} p_\m)\hem
    (\mathe^{\pounds_N^D} dx^\m)
    \hem=\hem
    p_\m\hem
    (\mathe^{\pounds_N^D} dx^\m)
    \,,
\end{align}
where
the last equality follows from
$\pounds_N^D p_\m = \i_N Dp_\m = 0$.
Therefore, from
Eqs.\,\oldeqref{eq:W.p}, 
\oldeqref{eq:pdz0}, 
and \oldeqref{eq:pdz-dist-cov},
it follows that
\begin{align}
    \label{eq:W.dz}
    W^\m{}_{\m'}\mem dz^{\m\sprime}
    \mem=\mem
        \mathe^{\pounds_N^D} dx^\m
    \,.
\end{align}

In the same way,
it can be shown that
the $\exp(\pounds_N^D)$
of a tensor-valued differential form
computes its value at the deviated point,
followed by the
parallel-transportation back to the original point
via dressing by the Wilson lines.
This fact will be referred to as
the \textit{dressing identity}.

\skip
\paragraph{Recursion for Jacobi Propagators}% 
Having set up the foundations of our formalism,
we now concern explicit evaluations.
In particular,
it follows from \eqref{eq:cov-magic} that
the right-hand side of \eqref{eq:W.dz} 
evaluates as
\begin{align}
    \label{eq:pdz1}
        dx^\m
	+
        Dy^\m
	+
        \sum_{\ell=2}^\infty 
            \frac{1}{\ell!}\mem
                (\hhnem{
                    (\i_ND)^{\ell-2} \i_N R^\m{}_\n
                }\hhem)\mem y^\n
    \,,
\end{align}
where $R^\m{}_\n$ denotes the Riemann curvature two-form such that $D^2 y^\m = R^\m{}_\n\mem y^\n$.
This computation is illustrated in Fig.\,\ref{fig:liecd-e}
as a curved deformation 
of the sequence of differential forms due to the Cartan magic formula.
As a tensor at the point $x$,
\eqref{eq:pdz1} eventually boils down to 
the following form:
\begin{align}
\begin{split}
    \label{eq:jacXY}
    W^\m{}_{\m'}\mem dz^{\m\sprime}
    \mem=\mem
    \mathe^{\pounds_N^D} dx^\m
   	\mem&=\mem
        X^\m{}_\s\mem dx^\s
        +
        Y^\m{}_\s\mem Dy^\s
    \,.
\end{split}
\end{align}
The objective now is to find the tensors
$X^\m{}_\s$ and $Y^\m{}_\s$ explicitly.
Note that
$W^{\m\sprime}{}_\m\mem X^\m{}_\s$
and
$W^{\m\sprime}{}_\m\mem Y^\m{}_\s$
are
exactly what are known as Jacobi propagators
(denoted as 
$K^{\m\sprime}{}_\s$ and $H^{\m\sprime}{}_\s$
in \rrcite{Vines:2014oba,dixon1979isolated}% 
),
where $W^{\m\sprime}{}_\m$ is the inverse of $W^\m{}_{\m'}$.
We find it preferable
to peel off the Wilson line,
for which the path-ordered exponential formula is well-known and amenable.
\begin{figure}
	{\begin{tikzpicture}
			\node[empty] (O) at (0,0) {};
			\node[empty] (X) at (3.0, 0) {};
			\node[empty] (Y) at (0, -0.9) {};
			\node[w] (a00) at ($(O)$) {$dx^\m$};
			\node[w] (a01) at ($(O)+1*(X)$) {$Dy^\m$};
			\node[w] (a02) at ($(O)+2.0*(X)$) {$0$};
			\node[w] (a10) at ($(O)+1*(Y)$) {$0$};
			\node[w] (a11) at ($(O)+1*(Y)+1*(X)$) {$(\i_NR^\m{}_\n)\hem y^\n$};
			\node[w] (a12) at ($(O)+1*(Y)+2*(X)$) {$0$};
			\node[w] (a21) at ($(O)+2*(Y)+1*(X)$) {$(\i_ND\mem \i_NR^\m{}_\n)\hem y^\n$};
			\node[w] (a22) at ($(O)+2*(Y)+2*(X)$) {$0$};
			\node[w] (a31) at ($(O)+3*(Y)+1*(X)$) {$\vdots$};
			\node[w] (phantom-a00) at ($(O)$) {};
			\node[w] (phantom-a01) at ($(O)+1*(X)$) {};
			\node[w] (phantom-a02) at ($(O)+2.0*(X)$) {};
			\node[w] (phantom-a10) at ($(O)+1*(Y)$) {};
			\node[w] (phantom-a11) at ($(O)+1*(Y)+1*(X)$) {};
			\node[w] (phantom-a12) at ($(O)+1*(Y)+2*(X)$) {};
			\node[w] (phantom-a10) at ($(O)+1*(Y)$) {};
			\node[w] (phantom-a11) at ($(O)+1*(Y)+1*(X)$) {};
			\node[w] (phantom-a12) at ($(O)+2*(Y)+2*(X)$) {};
			\node[w] (phantom-a21) at ($(O)+2*(Y)+1*(X)$) {};
			\node[w] (phantom-a31) at ($(O)+3*(Y)+1*(X)$) {};
			\draw[->] (a00)--(a01) node[midway,above] {\scriptsize $D\mem\i_N$};
			\draw[->] (a01)--(a02) node[] {};
			\draw[->] (a11)--(a12) node[] {};
			\draw[->] (a21)--(a22) node[] {};
			\draw[->] (phantom-a00)--(phantom-a10) node[midway,left] {\scriptsize $\i_ND$};
			\draw[->] (phantom-a01)--(phantom-a11) node[] {};
			\draw[->] (phantom-a11)--(phantom-a21) node[] {};
			\draw[->] (phantom-a21)--(phantom-a31) node[] {};
	\end{tikzpicture}}
	\caption{
		A sequence of one-forms originating from $dx^\m$
		via the covariant Lie derivative $\pounds_N^D = D\mem \i_N + \i_ND$.
		Note that $(D\i_N)(\i_ND) = 0$,
		while $D\hem dx^\m$
		vanishes for zero torsion.
	}
	\label{fig:liecd-e}
\end{figure}

To this end, one needs to compute
$(\i_ND)^{\ell-2} \i_NR^\m{}_\n$
for $\ell {\,\geq\,} 2$.
For $\ell {\,=\,} 3$, one finds
$\i_N D\mem \i_N R^\m{}_\n
=
	y^\k
	R^\m{}_{\n\r\s;\k}(x) $ $
		y^\r\mem dx^\s
	+
		R^\m{}_{\n\r\s}(x)\mem
			y^\r\hem Dy^\s
$.
When one hits this with a yet another $\i_ND$,
the covariant exterior derivative $D$ can
act on $Dy^\s$ to generate another Riemann tensor.
As a result,
one finds
two single-Riemann terms
and one double-Riemann term
at $\ell {\,=\,} 4$
(see Appendix\,\ref{organic-chemistry}).
In the same fashion, higher concatenations of Riemann tensors
arise at higher orders.

\begin{widetext}
A recursive structure can be identified 
in these calculations.
First of all, it follows that
\begin{align}
	\begin{split}
		\label{eq:boldQs}
		({\i_ND})^{\ell-2}\mem \i_N R^\m{}_\n\mem y^\n
		\mem=
		{}&{}
		\sum_{p=1}^{\lfloor{\ell/2}\rfloor}\kern-0.2em
			\sum_{\a \in \Omega_p(\ell)}\kern-0.2em
				\co{\a_1,\cdots,\a_p}\mem
				(Q_{\a_1} {\nem\cdots\mem} Q_{\a_p})^\m{}_\s
			\mem dx^\s
		+
		\sum_{p=1}^{\lfloor{\hnem\frac{\ell-1}{2}\hnem}\rfloor}\kern-0.3em
			\sum_{\a \in \Omega_p(\ell-1)}\kern-0.2em
				\cop{\a_1,\cdots,\a_p}\mem
				(Q_{\a_1} {\nem\cdots\mem} Q_{\a_p})^\m{}_\s
			\mem Dy^\s
		\,,
	\end{split}
\end{align}
where $\a$ runs over ordered partitions
such that
\begin{align}
	\a = (\a_1,\a_2,\cdots,\a_p)
		\,\in\,
		\Omega_p(\ell)
	\,\,\quad\iff\quad\,\,
	\a_1 + \a_2 + \cdots + \a_p 
	\mem=\mem \ell
	\,,\quad
	\a_i \geq 2
	\,.
\end{align}
Note that the total number of such partitions
at each $\ell$, i.e.,
$\sum_{p=1}^{\lfloor\ell/2\rfloor}\nem |\mem{ \Omega_p(\ell) }\hem|$,
is the $(\ell{\,-\mem}1)$\textsuperscript{th} Fibonacci number.
%Note that the total number of such partitions
%at each $\ell$ is the $(\ell{\,-\mem}1)$\textsuperscript{th} Fibonacci number.
%Note that the total number of such partitions
%at each $\ell$ is the $(\ell{\,-\mem}1)$\textsuperscript{th} Fibonacci number, i.e.,
%$\sum_{p=1}^{\lfloor\ell/2\rfloor}\nem |\mem{ \Omega_p(\ell) }\hem| = F_{\ell-1}$,
%where $F_1 = F_2 = 1$.
In \eqref{eq:boldQs},
the
``$Q$-tensors'' are defined as
\begin{align}
    \label{eq:def-Qten}
    (Q_\ell)^\m{}_\s
    {}:={}\mem&
	  y^{\k_1}{\cdots}y^{\k_\ell}\mem
	  R^\m{}_{\k_1\k_2\s;\k_3;\cdots;\k_\ell}\hnem(x)
    \qiq
    (Q_\ell)_{\m\s} = (Q_\ell)_{\s\m}
    \,,\quad
    (Q_\ell)^\m{}_\s\hem y^\s = 0
    \,.
\end{align}
The recursion relations
for the coefficients in \eqref{eq:boldQs}
are easily found from
identifying the following action of $\i_ND$:
\begin{subequations}
\begin{align}
	\i_ND
	\,\,\,:\,\,\,\,
		(Q_{\a_1} {\nem\cdots\mem} Q_{\a_p})^\m{}_\s
		\mem dx^\s
	&\,\,\mapsto\,\,
		\bbsq{\mem
			\sum_{i=1}^{p}\,
				(Q_{\a_1} {\nem\cdots\mem} Q_{\a_{i-1}} Q_{\a_i +1} Q_{\a_{i+1}} {\nem\cdots\mem} Q_{\a_p})^\m{}_\s
				\mem dx^\s
		}
		+
		(Q_{\a_1} {\nem\cdots\mem} Q_{\a_p})^\m{}_\s
			\mem Dy^\s
	\,,\\[-0.1\baselineskip]
		(Q_{\a_1} {\nem\cdots\mem} Q_{\a_p})^\m{}_\s
		\mem Dy^\s
	&\,\,\mapsto\,\,
		\bbsq{\mem
			\sum_{i=1}^{p}\,
				(Q_{\a_1} {\nem\cdots\mem} Q_{\a_{i-1}} Q_{\a_i +1} Q_{\a_{i+1}} {\nem\cdots\mem} Q_{\a_p})^\m{}_\s
				\mem Dy^\s
		}
		+
		(Q_{\a_1} {\nem\cdots\mem} Q_{\a_p} Q_2)^\m{}_\s
			\mem dx^\s
	\,.
\end{align}
\end{subequations}
With the proper identification of the boundary conditions \footnote{
	(a)
	$\co{\a_1,\a_2,\cdots,\a_p} = 0$,
	$\cop{\a_1,\a_2,\cdots,\a_p} = 0$
	if any of $\a_1,\a_2,\cdots,\a_p$ equals one.
	(b) $\co{2} = 1$, $\cop{2} = 0$.
},
the solution is determined as
\footnote{
	The generating function
	$G(\zeta_1,\cdots,\zeta_p) = \Sigma_\alpha c^{\alpha_1,\cdots,\alpha_p}\mem (\zeta_1)^{\alpha_1} \cdots (\zeta_p)^{\alpha_p}$
	is determined as
	$G(\zeta_1,\cdots,\zeta_p) = ((\zeta_p)^2/(\zeta_1 \mplus\cdots\mplus \zeta_p - 1)) \cdot (\zeta_1/(\zeta_1-1))^2 \cdot (\zeta_2/(\zeta_1\mplus\zeta_2 -1))^2 \cdots (\zeta_{p-1}/(\zeta_1\mplus\cdots\zeta_{p-1}-1))^2$.
}
\begin{align}
	\label{c-sols}
	c^{\a_1,\cdots,\a_p}
	\mem=\mem
		\prod_{i=1}^p
		\binom{
			\bigbig{
				\sum_{j=i}^p \a_j
			} \mminus 2
			\hem
		}{
			\a_i \mminus 2
		}
	\,,\quad
	c'^{\a_1,\cdots,\a_p}
	\mem=\mem
		\prod_{i=1}^p
		\binom{
			\bigbig{
				\sum_{j=i}^p \a_j
			} \mminus 1
			\hem
		}{
			\a_i \mminus 2
		}
	\,,
\end{align}
which describes products of binomial coefficients.
Finally, plugging in \eqref{c-sols} to \eqref{eq:boldQs},
we arrive at the following formula
for $\mathe^{\pounds_N^D} dx^\m$
from which
the tensors
$X$ and $Y$ in \eqref{eq:jacXY}
are readily read off:
\begin{align}
%\begin{split}
	\label{sol}
	\mathe^{\pounds_N^D} dx^\m \hnem= {}
&
	 	dx^\m
	 	+ Dy^\m
	 	\\[-0.2\baselineskip]
&
	 	+\mem \sum_{\ell=2}^\infty
	 		\frac{1}{\ell!}
			\sum_{p=1}^{\lfloor{\ell/2}\rfloor}\kern-0.2em
			\sum_{\a \in \Omega_p(\ell)}\kern-0.2em
	 			\bbsq{
	 				\prod_{i=1}^p
	 				\binom{
	 					\bigbig{
	 						\sum_{j=i}^p \a_j
	 					} \mminus 2
	 					\hem
	 				}{
	 					\a_i \mminus 2
	 				}
	 				\nem\nem
	 			}\mem
	 		\BB{
	 				(Q_{\a_1} {\nem\cdots\mem} Q_{\a_p})^\m{}_\s\, dx^\s
 				+ (\a_p\mminus2)\mem
	 				(Q_{\a_1} {\nem\cdots\mem} Q_{\a_{p-1}} Q_{\a_p - 1})^\m{}_\s\, Dy^\s
	 		}
	 \,.
	 \nonumber
%\end{split}
\end{align}
%\begin{align}
%	\label{sol}
%	X^\m{}_\s
%	\mem=\mem
%	\delta^\m{}_\s
%	 	+\mem \sum_{\ell=2}^\infty
%	 		\frac{1}{\ell!}
%			\sum_{p=1}^{\lfloor{\ell/2}\rfloor}\kern-0.2em
%			\sum_{\a \in \Omega_p(\ell)}\kern-0.2em
%	 				\prod_{i=1}^p
%	 				\binom{
%	 					\bigbig{
%	 						\sum_{j=i}^p \a_j
%	 					} \mminus 2
%	 					\hem
%	 				}{
%	 					\a_i \mminus 2
%	 				}
%		(Q_{\a_1} \cdots Q_{\a_p})^\m{}_\s
%	\,,\quad
%	Y^\m{}_\s
%	\mem=\mem
%	\delta^\m{}_\s
%	 	+\mem \sum_{\ell=2}^\infty
%	 		\frac{1}{\ell!}
%			\sum_{p=1}^{\lfloor{\ell/2}\rfloor}\kern-0.2em
%			\sum_{\a \in \Omega_p(\ell)}\kern-0.2em
%	 				\prod_{i=1}^p
%	 				\binom{
%	 					\bigbig{
%	 						\sum_{j=i}^p \a_j
%	 					} \mminus 2
%	 					\hem
%	 				}{
%	 					\a_i \mminus 2
%	 				}
%		(\a_p\mminus2)\mem
%		(Q_{\a_1} \cdots Q_{\a_{p-1}} Q_{\a_p - 1})^\m{}_\s
%	 \,.
%\end{align}
See
Appendix\:\ref{APP-XY}
for explicit enumerations up to $\O(y^{10})$.

\pagebreak
\end{widetext}

The explicit solution
for the Jacobi propagators
in \eqref{sol}
has not been spelled out in the literature
to our best knowledge,
though
Appendix 3 of Vines \cite{Vines:2014oba}
has identified a set of relevant recursion relations
by building upon
Ottewill and Wardell \cite{Ottewill:2009uj}
and Dixon \cite{dixon1979isolated}.

\skip
\paragraph{The Lagrangians}% 
We are now ready to derive
the all-orders GDE and its Lagrangian.
When described with the primed variables,
the first-order action of a free-falling test particle is
\begin{align}
	\label{eq:L0}
	\int\hhnem d\s\,
	\bigg[\,\,{
		p_{\m'}\hem \frac{d{z}^{\m\sprime}}{d\s}
		- \frac{e}{2}\mem \Big(\,{
			g^{\m\sprime\mem\n\sprime}\hhnem(z)\mem p_{\m'} p_{\n\sprime}
			+ m^2
		}\,\Big)
	}\mem\,\bigg]
	\,,
\end{align}
where $e$ is the einbein (a Lagrange multiplier),
and $m$ is the rest mass.
To describe
this particle from 
an observer's worldline, $\s \mapsto x^\m(\s)$,
we identify that
the first term
in \eqref{eq:L0}
originates from
the one-form $p_{\m'}\hem dz^{\m\sprime} = p_\m\mem (\hem {\exp(\pounds_N^D)\mem dx^\m} \hhem)$.
Consequently,
\eqref{eq:L0} boils down to
\begin{align}
	\label{eq:L1}
	\kern-0.4em
	\int\hhnem d\s\,
	\bigg[\,\,{
		p_\m\mem \bigg(\hem{
			X^\m{}_\n\hem 
			\frac{dx^\n}{d\s}
			{\mem+\,} Y^\m{}_\n\hem
			\frac{Dy^\n}{d\s}
		}\mem\bigg)\hnem
		- \frac{e}{2}\mem (p^2 {\mem+\,} m^2)
	}\mem\,\bigg]
	\,,
	\kern-0.5em
\end{align}
where $p^2 = g^{\m\n}\hhnem(x)\mem p_\m\hem p_\n$
since
Wilson lines due to the Levi-Civita connection respect the metric.
By integrating out $p_\m$,
we also find
a second-order Lagrangian:
\begin{align}
	\label{eq:L2}
	\int\hhnem d\s\,
	\bigg[\,\,{
		\frac{1}{2e}\mem
		\bigg(\mem{
			X\mem \frac{dx}{d\s}
			{\mem+\,} 
			Y\mem \frac{Dy}{d\s}
		}\mem\bigg)^{\nem2}
		- \frac{m^2e}{2}
	}\mem\,\bigg]
	\,.
\end{align}
Adopting an invariant measure of time $d\t = me\mem d\s$
for $m\neq0$
reproduces Vines \cite{Vines:2014oba}'s action
for affinely parameterized worldlines (isochronous correspondence \cite{aleksandrov1979geodesic,Vines:2014oba}):
\begin{align}
	\label{eq:L2-tau}
	m\mem
	\int\hnem d\t\,\,
	\frac{1}{2}\mem
	\bigg(\mem{
		\bigbig{\nem
			X\mem u {\mem+\,} Y\mem v
		}^{\nem2}
		- 1
	}\,\bigg)
	\,.
\end{align}
Here, we have denoted $u^\m := dx^\m/d\t$ and $v^\m := Dy^\m/d\t$.

Eqs.\,(\ref{eq:L1})-(\ref{eq:L2-tau})
provide exact first-order and second-order Lagrangian formulations of the all-orders geodesic deviation,
where the deviation $y^\m$ is defined as a vector attached to the observer's worldline.
We clarify that the observer's worldline is introduced as a nondynamical reference \cite{Vines:2014oba},
while $y^\m$ and $p_\m$ are dynamical variables \footnote{
	It might be interesting to understand
	the transformations to other correspondences 
	as gauge transformations in the first-order setup.
}.

The Lagrangian in \eqref{eq:L2-tau}
is explicitly given
up to $\O(y^{10})$
in Appendix~\ref{APP-L2},
showing perfect agreement with
the previous $\O(y^5)$ result
due to 
Vines \cite{Vines:2014oba}.

Note that the second-order actions in \eqrefs{eq:L2}{eq:L2-tau}
could have been directly obtained
by implementing our formalism simply in the tangent bundle $T\M$,
instead of employing the extended bundle $\P$.

\skip
\paragraph{The All-Orders GDE}% 
The all-orders GDE follows by varying the above Lagrangians.
Otherwise, it can also be derived 
from our formalism
in the following way.

Consider 
the first-order formulation of the free-falling equations of motion,
associated with \eqref{eq:L0}:
\begin{align}
	\label{ff'}
	\frac{dz^{\m\sprime}}{d\s}
	\,=\,
		e\mem p^{\m\sprime}
	\,,\quad
	\frac{Dp^{\m\sprime}}{d\s}
	\,=\,
		0
	\,.
\end{align}
The idea is to re-covariantize \eqref{ff'}
at the observer's position, $x^\m$,
via the geodesic Wilson line dressing.
Earlier, we have obtained
$W^\m{}_{\m'}\hem dz^{\m\sprime} = X^\m{}_\n\mem dx^\n + Y^\m{}_\n Dy^\n$ 
in \eqref{eq:jacXY}.
This applies to the 
first equation in \eqref{ff'}.
Taking a similar approach for the second equation as well,
we obtain a first-order formulation of the all-orders GDE:\hnem
\begin{subequations}
\label{1GDE}
\begin{align}
	\label{1GDE.x}
	e\mem p^\m
	\,&=\,
		X^\m{}_\n\mem \frac{dx^\n}{d\s}
		+ Y^\m{}_\n\mem \frac{Dy^\n}{d\s}
	\,,\\
	\label{1GDE.p}
	\frac{Dp^\m}{d\s}
	\,&=\,
		-\bb{
			\acute{X}^\m{}_{\n\s}\mem \frac{dx^\s}{d\s}
			+ \acute{Y}^\m{}_{\n\s}\mem \frac{Dy^\s}{d\s}
		}\mem p^\n
	\,.
\end{align}
\end{subequations}

\begin{widetext}
For obtaining \eqref{1GDE.p}, we have 
applied the dressing identity to the one-form $Dp^\m$:
\begin{align}
\begin{split}
	\label{dressing-Dp}
	W^\m{}_{\m'}
	Dp^{\m'}
	\,&=\,
		\mathe^{\pounds_N^D} Dp^\m
	\,=\,
		Dp^\m
		+
			\acute{X}^\m{}_{\n\s}\mem p^\n\mem dx^\s
		+
			\acute{Y}^\m{}_{\n\s}\mem p^\n\mem Dy^\s
	\,.
\end{split}
\end{align}
The tensors $\acute{X}^\m{}_{\n\s}$ and $\acute{Y}^\m{}_{\n\s}$ 
are given as
\begin{subequations}
	\label{sol-acute}
\begin{align}
	\label{sol-acute.X}
		\acute{X}^\m{}_{\n\s}
		\mem&=\mem
	 	\sum_{\ell=2}^\infty
	 		\frac{1}{(\ell-1)!}
			\sum_{p=1}^{\lfloor{\ell/2}\rfloor}\kern-0.2em
			\sum_{\a \in \Omega_p(\ell)}\kern-0.2em
	 			\bbsq{
	 				\prod_{i=1}^p
	 				\binom{
	 					\bigbig{
	 						\sum_{j=i}^p \a_j
	 					} \mminus 2
	 					\hem
	 				}{
	 					\a_i \mminus 2
	 				}
	 				\nem\nem
	 			}\,
	 				(\acute{Q}_{\a_1})^\m{}_{\n\k}
	 				\mem
	 				(Q_{\a_2} \cdots Q_{\a_p})^\k{}_\s
	 \,,\\
 	\label{sol-acute.Y}
 		\acute{Y}^\m{}_{\n\s}
 		\mem&=\mem
	 	\sum_{\ell=2}^\infty
	 		\frac{1}{(\ell-1)!}	
			\sum_{p=1}^{\lfloor{\ell/2}\rfloor}\kern-0.2em
			\sum_{\a \in \Omega_p(\ell)}\kern-0.2em
	 			\bbsq{
	 				\prod_{i=1}^p
	 				\binom{
	 					\bigbig{
	 						\sum_{j=i}^p \a_j
	 					} \mminus 2
	 					\hem
	 				}{
	 					\a_i \mminus 2
	 				}
	 				\nem\nem
	 			}\,
	 				(\a_p\mminus2)\mem
	 				(\acute{Q}_{\a_1})^\m{}_{\n\k}
	 				\mem
	 				(Q_{\a_2} \cdots Q_{\a_{p-1}} Q_{\a_p - 1})^\k{}_\s
	\,,
\end{align}
\end{subequations}
which simply replaces $\ell!$ in \eqref{sol}
to $(\ell{\mem-\,}1)!$
regarding the combinatorial factors.
Here, we have defined, for $\ell \geq 2$,
\begin{align}
	(\acute{Q}_\ell)^\m{}_{\n\s}
	:= 
	    y^{\k_1}{\cdots}y^{\k_{\ell-1}}
	    R^\m{}_{\n\k_1\s;\k_2\cdots\k_{\ell-1}}\hnem(x)
	\qiq
\nem\left\{\,
\begin{aligned}[c]
	&
	(\acute{Q}_\ell)_{\m\n\s} = -(\acute{Q}_\ell)_{\n\m\s}
	\,,\quad
	\\
	&
	(\acute{Q}_\ell)_\wrap{[\m\n\s]} = 0
	\,,\quad
\end{aligned}
\begin{aligned}[c]
	&
	(\acute{Q}_\ell)_{\m\n\s} y^\s = 0
	\,,\\
	&
	(\acute{Q}_\ell)^\m{}_{\n\s}\mem y^\n = (Q_\ell)^\m{}_\s
	\,.
\end{aligned}
\right.
\end{align}

Combining \eqrefs{1GDE.x}{1GDE.p},
the second-order formulation
of the all-orders GDE 
in the isochronous correspondence
is found as
\begin{align}
\begin{split}
	\label{2GDE}
	- Y^\m{}_\n\mem \frac{Dv^\n}{d\tau}
	\mem=\mem
		\BB{				
			\nabla_\r X^\m{}_\s
			{\,+\,} \acute{X}^\m{}_{\n\r}\mem X^\n{}_\s
		\nem}\mem u^\r u^\s
		+ 2\mem
		\BB{					
			\nabla_\r Y^\m{}_\s
			{\,+\,} \acute{X}^\m{}_{\n\r}\mem Y^\n{}_\s
		\nem}\mem u^\r v^\s
		+
		\bb{
			\frac{\partial}{\partial y^\r}\mem Y^\m{}_\s
			{\,+\,} \acute{Y}^\m{}_{\n\r}\mem Y^\n{}_\s
		\nem}\mem v^\r v^\s
	\,,
\end{split}
\end{align}
where
it should be understood that $\nabla_\r$ will be acted only on the Riemann tensors inside the $Q$-tensors.

\pagebreak

By inverting the matrix $Y^\m{}_\n$ on the left-hand side,
the explicit GDE in the form $-Dv^\m \nem/d\t = \cdots$
is obtained 
from \eqref{2GDE}
and shown
up to $\O(y^{10})$
in the ancillary file \texttt{All.nb}.
See Appendix~\ref{APP-2GDE} for the specifics of this computation.
Moreover,
we have also verified that 
the GDE obtained in this way
is exactly reproduced from varying the second-order Lagrangian in \eqref{eq:L2-tau},
up to $\O(y^5)$
in the ancillary file \texttt{Low.nb}.
Here, we spell out the explicit GDE up to $\O(y^5)$ \footnote{
	The $\O(y^4)$ GDE provided by Vines \cite{Vines:2014oba}
	describes
	$-1$ (instead of $4$) for the coefficient of the term
	\smash{$\protect\acQ_2(Q_3u,u)$}
	and
	$-8$ (instead of $8$) for the coefficient of the terms
	\smash{$\protect\acQ_2(v,Q_2u)$} and \smash{$\protect\acQ_2(Q_2v,u)$}.
	As our $\O(y^5)$ Lagrangian in \oldeqref{eq:GDEL-5} agrees flawlessly with Eq.\,(5) in Vines,
	and since it is explicitly verified in the ancillary file \texttt{Low.nb}
	that the variation the $\O(y^5)$ Lagrangian
	implies our $\O(y^4)$ GDE,
	we suppose the above mismatched coefficients are typos.
}:
\begin{align}
	\label{GDEX5}
	&
	{- \frac{Dv}{d\t}}
	{}={}
%	&
	\acQ_2(u,u)
	+
	\frac{1}{2!}\mem
	\bbsq{
		\BB{
			\acQ_3(u,u)
			+
			\nabla_{\nem u} (Q_2u)
		}
		+ 4\mem \acQ_2(v,u)
	\mem}
	\\\nonumber
	&
	+ \frac{1}{3!}
	\lrsq{
		\BB{
			\acQ_4(u,u)
			+
			\nabla_{\nem u} (Q_3u)
			+
			\acQ_2(u,Q_2u)
			+
			3\mem \acQ_2(Q_2u,u)
		\hls{
			-
			Q_2\mem \acQ_2(u,u)
		}
		}
		+
		\BB{
			2\mem \nabla_{\nem v} (Q_2u)
			+ 6\mem \acQ_3(v,u)
		}
		+
			4\mem \acQ_2(v,v)
	}
	\\\nonumber
	&
	+ \frac{1}{4!} \lrsq{\begin{aligned}[c]
		&
		\lrp{
		\begin{aligned}[c]
		&
			\acQ_5(u,u)
			+
			\nabla_{\nem u} (Q_4u)
			+
			(\nabla_{\nem u} Q_2)\mem (Q_2u)
			+
			Q_2\mem \nabla_{\nem u} (Q_2u)
		\hls{
			- 2\mem Q_2\mem \nabla_{\nem u} (Q_2u)
		}
		\\
		&
			+
			6\mem \acQ_3(Q_2u,u)
			+
			3\mem \acQ_3(u,Q_2u)
			+
			\acQ_2(u,Q_3u)
			+
			4\mem \acQ_2(Q_3u,u)
		\hls{
			- 2\mem Q_2\mem \acQ_3(u,u)
			- 2\mem Q_3\mem \acQ_2(u,u)
		}
		\end{aligned}
		}
		\\
		&
		+
		\BB{
			4\mem \nabla_{\nem u} (Q_3v)
			+ 8\mem \acQ_4(v,u)
			+ 8\mem \acQ_2(v,Q_2u)
			+ 8\mem \acQ_2(Q_2v,u)
		\hls{
			- 8\mem Q_2\mem \acQ_2(v,u)
		}
		}
		\\
		&
		+
		\BB{
			2\mem \nabla_{\nem v} (Q_2v)
			+
			10\mem \acQ_3(v,v)
		}
	\end{aligned}}
	\\\nonumber
	&
	+ \frac{1}{5!} \lrsq{\begin{aligned}[c]
		&
		\lrp{\,
		\begin{aligned}[c]
		&
			\acQ_6(u,u)
			+
			\nabla_{\nem u} (Q_5u)
			+
			10\mem \acQ_4(Q_2u,u)
			+
			10\mem \acQ_3(Q_3u,u)
			+
			5\mem \acQ_2(Q_4u,u)
			+
			5\mem \acQ_2(Q_2Q_2u,u)
		\\
		&
			+
			Q_2\mem \nabla_{\nem u} (Q_3u)
			+
			3\mem Q_3\mem \nabla_{\nem u} (Q_2u)
			+
			(\nabla_{\nem u} Q_2)\mem Q_3
			+
			3\mem (\nabla_{\nem u} Q_3)\mem Q_2
		\hls{
			- \tfrac{10}{3}\mem Q_2\mem \nabla_{\nem u} (Q_3u)
			- 5\mem Q_3\mem \nabla_{\nem u} (Q_2u)
		}
		\\
		&
			+
			\acQ_2(u,Q_4u)
			+
			4\mem \acQ_3(u,Q_3u)
			+
			6\mem \acQ_4(u,Q_2u)
			+
			10\mem \acQ_2(Q_2u,Q_2u)
			+
			\acQ_2(u,Q_2Q_2u)
		\hls{
			- \tfrac{10}{3}\mem Q_2\mem \acQ_4(u,u)
		}
		\\
		&
		\hls{
			- 10\mem Q_2\mem \acQ_2(Q_2u,u)
			- 5\mem Q_3\mem \acQ_3(u,u)
			- 3\mem Q_4\mem \acQ_2(u,u)
			+ \tfrac{7}{3}\mem Q_2 Q_2\mem \acQ_2(u,u)
			- \tfrac{10}{3}\mem Q_2 \acQ_2(u,Q_2u)
		}
		\end{aligned}
		}
		\\\nonumber
		&
		+
		\lrp{
		\begin{aligned}[c]
		&
			10\mem \acQ_5(v,u)
			+
			6\mem \nabla_{\nem u} (Q_4v)
			+
			2\mem Q_2\mem \nabla_{\nem u} (Q_2v)
			+
			2\mem (\nabla_{\nem u} Q_2)\mem (Q_2v)
		\hls{
			- \tfrac{20}{3}\mem Q_2\mem \nabla_{\nem u} (Q_2v)
		}
			+
			20\mem \acQ_3(Q_2v,u)
		\\
		&
			+
			20\mem \acQ_2(Q_3v,u)
			+
			10\mem \acQ_2(v,Q_3u)
			+
			30\mem \acQ_3(v,Q_2u)
		\hls{
			-20\mem Q_2\mem \acQ_3(v,u)
			-20\mem Q_3\mem \acQ_2(v,u)
		}
		\end{aligned}
		}
		\\\nonumber
		&
		+
		\BB{
			6\mem \nabla_{\nem v} (Q_3v)
			+
			18\mem \acQ_4(v,v)
			+
			6\mem \acQ_2(Q_2v,v)
			- 2\mem Q_2\mem \acQ_2(v,v)
			+ 10\mem \acQ_2(v,Q_2v)
		\hls{
			- \tfrac{40}{3}\mem Q_2\mem \acQ_2(v,v)
		}
		}
	\end{aligned}}
	\\\nonumber
	&
	+ \O(y^6)
	\,.
\end{align}
We have
color-coded terms due to $Y^{-1}$
and
adopted a condensed notation:
$Dv/d\t \to Dv^\m \nem/d\t$,
$\acQ_3(v,u) \to (\smash{\acQ_3})^\m{}_{\r\s} $ $  v^\r u^\s$,
$Q_2\mem \smash{\acQ_2}(u,u) \to (Q_2)^\m{}_\n\mem (\smash{\acQ_2})^\n{}_{\r\s}\mem u^\r u^\s$,
$\nabla_{\nem v} (Q_2u) \to v^\r\mem (\nabla_{\nem \r} (Q_2)^\m{}_\s)\mem u^\s$,
$(\nabla_{\nem u} Q_2)\hem (Q_2u) \to u^\r\mem (\nabla_{\nem \r}(Q_2)^\m{}_\s)\mem (Q_2)^\s{}_\k\mem u^\k$,
etc.
The sources of
the minus signs on the right-hand side
are
either $Y^{-1}$ or 
the Riemann tensor identities
used for simplifying the quadratic-in-$v$ part.
\end{widetext}

\skip
\paragraph{Zero-Torsion Identities}% 
One may notice that 
the behavior of
\smash{$\acX^\m{}_{\n\r}$} and \smash{$\acY^\m{}_{\n\r}$}
in \eqref{2GDE}
is suggestive of 
connection coefficients.
In fact, 
they arise from
the conjugation
$
	\smash{\mathe^{\pounds^D_N}\nem D\mem \mathe^{-\pounds^D_N}}
	=
	\smash{\mathe^{[\pounds^D_N,\blank]}\nem D}
$
of the covariant exterior derivative.
In turn,
it can be seen that
they encode the covariant derivative
at the deviated point:
see \eqref{avatar-gravity}.

On a related note,
the torsion-free condition 
for the Levi-Civita connection,
as $Ddx^\m = \Gamma^\m{}_{\r\s}\mem dx^\s \swedge dx^\r = 0$,
implies
$0 
= \smash{
	(\mathe^{\pounds_N^D} D\mem \mathe^{-\pounds_N^D})
	(\mathe^{\pounds_N^D} dx^\m) 
}
= 
	D(X^\m{}_\s dx^\s {+\hem} $ $ Y^\m{}_\s Dy^\s)
	+ 
		(\smash{\acX}^\m{}_{\n\r} dx^\r {\hem+\hem} \smash{\acY}^\m{}_{\n\r} Dy^\r)
		\wedge 
		(X^\n{}_\s dx^\s {+\hem} Y^\n{}_\s Dy^\s)
$,
which unpacks into three identities:
\begin{subequations}
\label{torid}
\begin{align}
	\label{torid-a}
	\nabla_\wrap{[\r} X^\m{}_\wrap{\s]}
	{\mem+\mem} \acX^\m{}_\wrap{\n[\r}\mem X^\n{}_\wrap{\s]}
	\mem&=\mem
		- \tfrac{1}{2}\mem Y^\m{}_\l\mem R^\l{}_{\n\r\s}
	\,,\\
	\label{torid-b}
	\nabla_\r Y^\m{}_\s
		{\mem+\mem} \acute{X}^\m{}_{\n\r}\mem Y^\n{}_\s
	\mem&=\mem
	\frac{\partial}{\partial y^\r}\mem X^\m{}_\s
		{\mem+\mem} \acute{Y}^\m{}_{\n\r}\mem X^\n{}_\s
	\,,\\
	\label{torid-c}
	\frac{\partial}{\partial y^{[\r}}\mem Y^\m{}_\wrap{\s]}
	{\mem+\mem} \acY^\m{}_\wrap{\n[\r}\mem Y^\n{}_\wrap{\s]}
	\mem&=\mem
		0
	\,.
\end{align}
\end{subequations}
Especially,
we have made a use of \eqref{torid-b}
in \eqref{2GDE}
to simplify the computation of the term linear in both $u$ and $v$.
The ancillary file \texttt{Low.nb}
provides an explicit check of Eqs.\,\oldeqref{torid-a}-\oldeqref{torid-c} up to $\O(y^5)$ or $\O(y^4)$,
which exploits various identities about the Riemann tensor.
Also, note that more identities follow 
in a similar fashion
by conjugating $D^2 = R$, $[D, R] = 0$, etc.

\skip
\paragraph{Summary and Outlook}% 
In this paper, we revisited the problem of finding the all-orders-exact GDE for finite separations
by formulating
geodesic deviation and transport
as a flow along a vector field in tangent bundle.
The technique of covariant Lie derivative
then systematically defines and generates
various bitensors with the parallel propagators peeled off,
directly producing
manifestly covariantized expressions at the original point.
This achieves an in-in formalism for geodesic deviation
that serves as an alternative to the Synge calculus.

The explicit outcomes are the all-orders formula for the Jacobi propagators in \eqref{sol}
as well as
the second-order GDE and its Lagrangian
given up to $\O(y^{10})$.

Our framework is versatile and
could find further applications
in the context of 
quantum field theory:
manifestly covariant perturbation theories 
for gauge and gravitational interactions,
worldline formalism \cite{Feynman:1948ur,Bern:1990cu,Strassler:1992zr,Schubert:2001he,schwartz2014qft},
or field space geometry \cite{Cheung:2021yog,Assi:2023zid}
and sigma models \cite{Alvarez-Gaume:1980zra,Alvarez-Gaume:1981exa,Callan:1985ia}.
Appendix~\ref{gaugetheory} 
implements our formalism in nonabelian gauge theories,
which describes gauge-covariant translations.

Moreover, we realize that
our innocuous attempt to remaster the fundamental subject of 
all-orders-in-deviation GDE
%this classic subject of the all-orders-in-deviation GDE
surprisingly connects to 
a persistent and seemingly disparate problem in 
% 
%modern gravitational physics:
the modern times:
finding the all-orders-in-spin equations of motion
for the Kerr black hole 
in its effective point-particle description
\cite{Levi:2015msa,Levi:2018nxp,Porto:2016pyg,Guevara:2018wpp,Guevara:2019fsj,chkl2019,aho2020,gmoov}.
By implementing a probe counterpart of the 
Newman-Janis algorithm \cite{Newman:1965tw-janis},
we have found that 
the all-orders GDE for an imaginary deviation
can deduce a part of the black hole's
all-orders-in-spin equations of motion,
yielding a nonlinear completion of the 
Mathisson-Papapetrou-Dixon  \cite{Mathisson:1937zz,Papapetrou:1951pa,Dixon:1970zza}
equations.
More details will be presented in a follow-up article
\cite{probe-nj}.

\medskip
\paragraph{Acknowledgements}%
	We acknowledge the use of the xAct package \cite{xTensor}
	in Mathematica.
	J.-H.K. is supported by
	the Department of Energy (Grant No. DE-SC0011632) and by the Walter Burke Institute for Theoretical Physics. 
\pagebreak
\appendix
\onecolumngrid
	\section{The ``Organic Chemistry'' of Covariant Lie Derivative Calculus}% 
\label{organic-chemistry}

In this appendix, we devise
a slight extension of
the Penrose graphical notation \cite{penrose1956tensor,penrose1971applications,penr04-tensor}
with the following rules.
\begin{align}
	\label{chem-legend}
	\adjustbox{valign=c}{\includegraphics[scale=0.3,
		trim={0 120pt 0 0},clip
	]{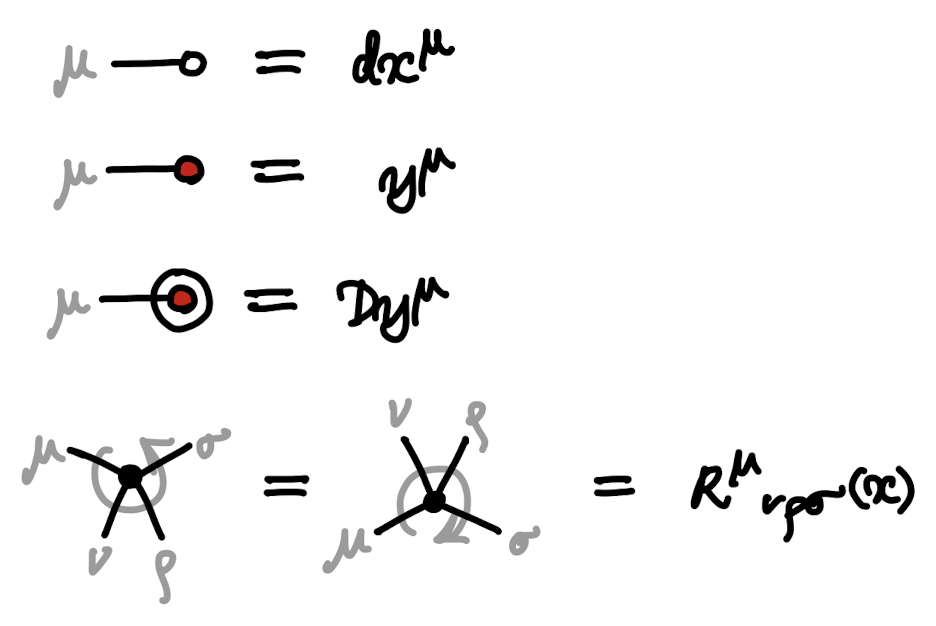}}
	\kern-4em
	\adjustbox{valign=c}{\includegraphics[scale=0.3,
		trim={0 0 0 200pt},clip
	]{figs/legend}}
\end{align}
The higher covariant derivatives of the Riemann tensor will be denoted as the following.
\begin{align}
	\label{chem-carbon}
	\adjustbox{valign=c}{\includegraphics[scale=0.3]{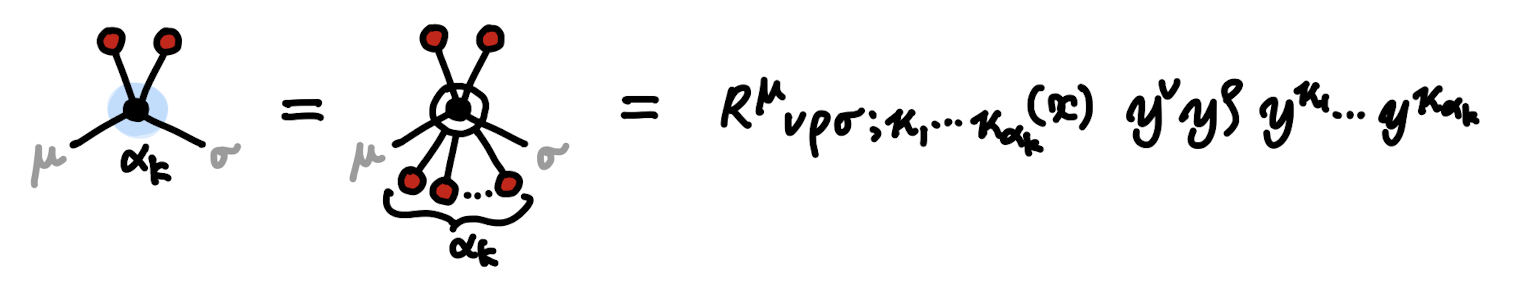}}
\end{align}
The relevant ``molecular backbones'' are the ``carbon'' (as Riemann tensor) chains of the following form.
\begin{align}
	\label{chem-chain}
	\adjustbox{valign=c}{\includegraphics[scale=0.3]{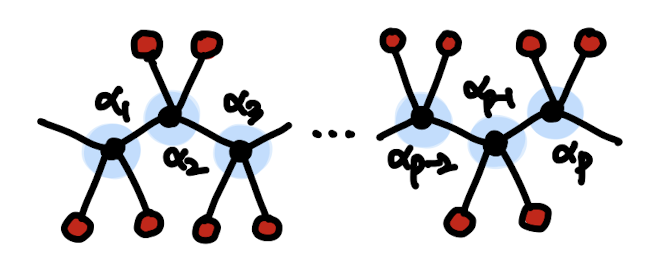}}
\end{align}
The computation of
$\i_N D\mem (\i_N R^\m{}_\n)$
and
$(\i_N D)^2\mem (\i_N R^\m{}_\n)$
proceeds as the following.
\begin{align}
	\label{chem-reaction}
	\adjustbox{valign=c}{\includegraphics[scale=0.3]{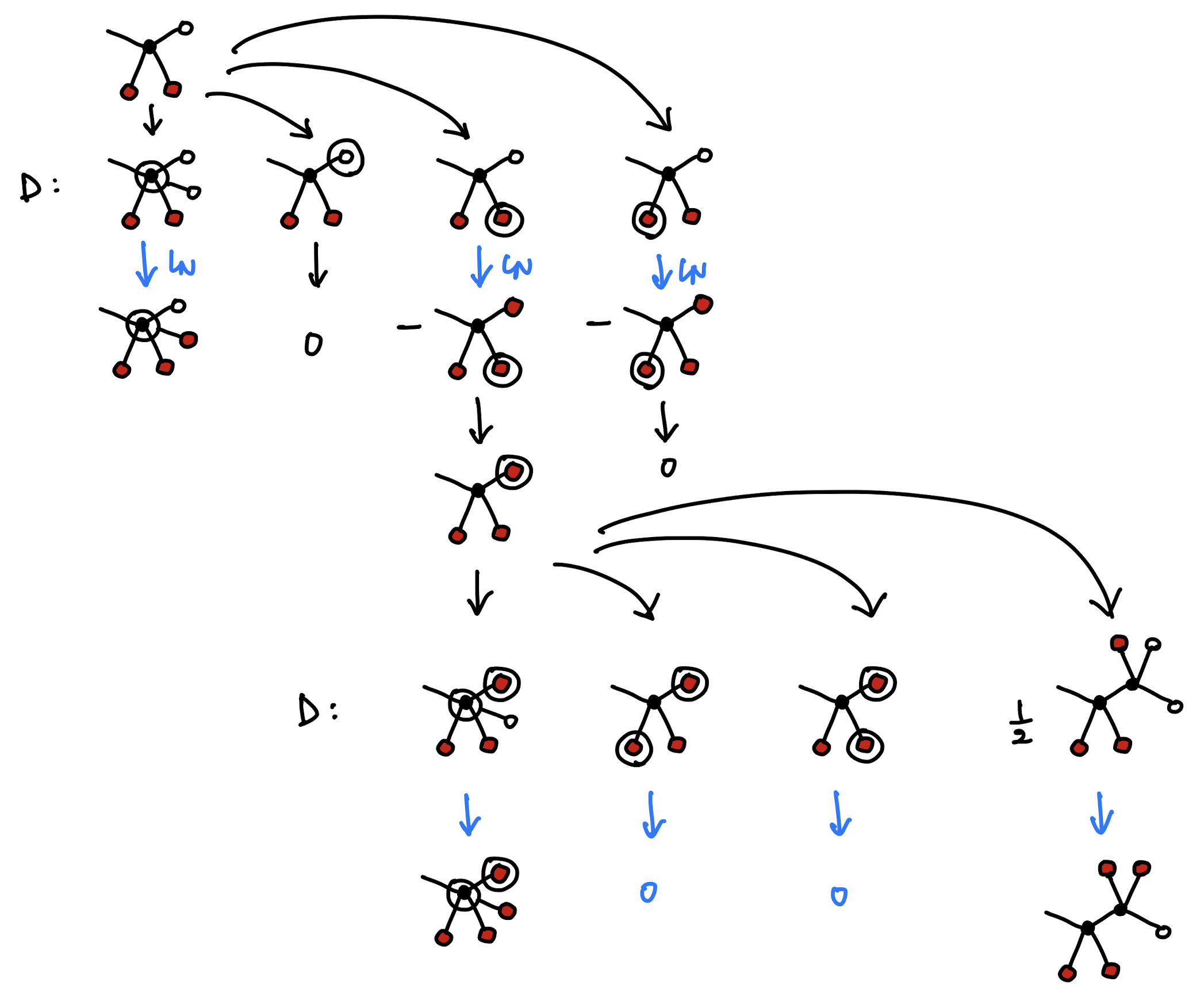}}
\end{align}
The ordering for the differential forms should be clear
from the orientations designated in \eqref{chem-legend}.
The interior product $\i_N$ 
is a ``reduction reaction''
that substitutes a ``hyrdoxyl group'' (as $dx$) with a ``hydrogen'' (as $y$).
Observe the mechanism for the ``carbon polymerization'' (as $Q$-tensor concatenation).

\pagebreak

\section{Explicit Results up to Tenth Order}% 
\label{10th}

\subsection{The Jacobi Propagators}% 
\label{APP-XY}% 
The Jacobi propagators 
with the parallel propagator peeled off
are given in \eqref{sol}.
It is then trivial to enumerate them explicitly up to any desired order,
say $\O(y^{10})$.
With
the definition of the $Q$-tensors
in \eqref{eq:def-Qten},
they are
\begin{align}
    \label{eq:Qaction-1}
	Dy^\m
	= {}\phantom{+}{}&
		Dy^\m
	\,,\\
    \label{eq:Qaction-2}
    (\i_N R^\m{}_\n)\hem y^\n
    = {}\phantom{+}{}&
        (Q_2)^\m{}_\s \mem{dx^\s}
    \,,\\
    \label{eq:Qaction-3}
    ({\i_ND}\mem \i_N R^\m{}_\n)\hem y^\n
    = {}\phantom{+}{}&
        (Q_3)^\m{}_\s \mem{dx^\s}
        + (Q_2)^\m{}_\s \mem{Dy^\s}
    \,,\\
    \label{eq:Qaction-4}
    (\hhnem{({\i_ND})^2\mem \i_N R^\m{}_\n}\hhhem)\hem y^\n
    = {}\phantom{+}{}&
        (Q_4 + Q_2 Q_2)^\m{}_\s \mem{dx^\s}
        + (2 Q_3)^\m{}_\s \mem{Dy^\s}
    \,,\\
    \label{eq:Qaction-5}
    (\hhnem{({\i_ND})^3\mem \i_N R^\m{}_\n}\hhhem)\hem y^\n
    = {}\phantom{+}{}&
        (Q_5 + 3 Q_3 Q_2 + Q_2 Q_3)^\m{}_\s \mem{dx^\s}
        + (3 Q_4 + Q_2 Q_2)^\m{}_\s \mem{Dy^\s}
    \,,\\
    \label{eq:Qaction-6}
    (\hhnem{({\i_ND})^4\mem \i_N R^\m{}_\n}\hhhem)\hem y^\n
    = {}\phantom{+}{}&
        (Q_6 + 6 Q_4 Q_2 + 4 Q_3 Q_3 + \phantom{2}Q_2 Q_4)^\m{}_\s \mem{dx^\s}
    \nonumber\\ {}+{}&
        (4 Q_5 \phantom{{}+Q_4Q_2{}} + 4 Q_3 Q_2 + 2 Q_2 Q_3)^\m{}_\s \mem{Dy^\s}
    \nonumber\\ {}+{}&
        (Q_2 Q_2 Q_2)^\m{}_\s \mem{dx^\s}
    \,,\\
    \label{eq:Qaction-7}
    (\hhnem{({\i_ND})^5\mem \i_N R^\m{}_\n}\hhhem)\hem y^\n
    = {}\phantom{+}{}&
        (Q_7 
            + 10 Q_5 Q_2 + 10 Q_4 Q_3 + \phantom{1}5 Q_3 Q_4 + \phantom{3}Q_2 Q_5
        )^\m{}_\s \mem{dx^\s}
    \nonumber\\ {}+{}&
        (5 Q_6 \phantom{{}+1Q_5Q_2{}} 
            + 10 Q_4 Q_2 + 10 Q_3 Q_3 + 3 Q_2 Q_4
        )^\m{}_\s \mem{Dy^\s}
    \nonumber\\ {}+{}&
        (
            5Q_3 Q_2 Q_2
            + 3Q_2 Q_3 Q_2
            + Q_2 Q_2 Q_3
        )^\m{}_\s \mem{dx^\s}
    \nonumber\\ {}+{}&
        (Q_2 Q_2 Q_2)^\m{}_\s \mem{Dy^\s}
    \,,\\
    \label{eq:Qaction-8}
    (\hhnem{({\i_ND})^6\mem \i_N R^\m{}_\n}\hhhem)\hem y^\n
    = {}\phantom{+}{}&
        (Q_8
            + 15 Q_6 Q_2 + 20 Q_5 Q_3 + 15 Q_4 Q_4 + \phantom{1}6 Q_3 Q_5 + \phantom{4}Q_2 Q_6
        )^\m{}_\s \mem{dx^\s}
    \nonumber\\ {}+{}&
        (6Q_7 \phantom{{}+1Q_5Q_2{}} 
            + 20 Q_5 Q_2 + 30 Q_4 Q_3 + 18 Q_3 Q_4 + 4 Q_2 Q_5
        )^\m{}_\s \mem{Dy^\s}
    \nonumber\\ {}+{}&
        \bigg(
        \begin{array}{l}
            \phantom{+}{}
            15 Q_4 Q_2 Q_2
            + 18 Q_3 Q_3 Q_2
            + 6 Q_2 Q_4 Q_2
            \\
            + \phantom{00}\mathllap{6} Q_3 Q_2 Q_3
            + \phantom{00}\mathllap{4}  Q_2 Q_3 Q_3 
            + \phantom{0} Q_2 Q_2 Q_4
        \end{array}
        \bigg){\vphantom{\Big)}}^\m{\vphantom{\Big)}}_\s \,{dx^\s}
    \nonumber\\ {}+{}&
        (
            6 Q_3 Q_2 Q_2 
            + 4 Q_2 Q_3 Q_2
            + 2 Q_2 Q_2 Q_3
        )^\m{}_\s \mem{Dy^\s}
    \nonumber\\ {}+{}&
        (Q_2 Q_2 Q_2 Q_2)^\m{}_\s \mem{dx^\s}
    \,,\\
    \label{eq:Qaction-9}
    (\hhnem{({\i_ND})^7\mem \i_N R^\m{}_\n}\hhhem)\hem y^\n
    = {}\phantom{+}{}&
        (Q_9
            + 21 Q_7 Q_2 + 35 Q_6 Q_3 + 35 Q_5 Q_4 + 21 Q_4 Q_5 + \phantom{2}7 Q_3 Q_6 + \phantom{5}Q_2 Q_7
        )^\m{}_\s \mem{dx^\s}
    \nonumber\\ {}+{}&
        (7Q_8 \phantom{{}+1Q_5Q_2{}} 
            + 35 Q_6 Q_2 + 70 Q_5 Q_3 + 63 Q_4 Q_4 + 28 Q_3 Q_5 + 5 Q_2 Q_6
        )^\m{}_\s \mem{Dy^\s}
    \nonumber\\ {}+{}&
        \left(
        \begin{array}{l}
            \phantom{+}{} 35 Q_5 Q_2 Q_2 
            + 63 Q_4 Q_3 Q_2 
            + 42 Q_3 Q_4 Q_2
            + 10 Q_2 Q_5 Q_2
            \\
            + 21 Q_4 Q_2 Q_3
            + 28 Q_3 Q_3 Q_3
            + 10 Q_2 Q_4 Q_3
            \\
            + \phantom{00}\mathllap{7} Q_3 Q_2 Q_4
            + \phantom{00}\mathllap{5} Q_2 Q_3 Q_4
            + \phantom{00} Q_2 Q_2 Q_5
        \end{array}
        \right){\vphantom{\Big)}}\kern-0.2em{\vphantom{\Big)}}^\m{\vphantom{\Big)}}_\s \,{dx^\s}
    \nonumber\\ {}+{}&
        \bigg(
        \begin{array}{l}
            \phantom{+}{} 
            21 Q_4 Q_2 Q_2 
            + 28 Q_3 Q_3 Q_2
            + 10 Q_2 Q_4 Q_2 
            \\
            + 14 Q_3 Q_2 Q_3
            + 10 Q_2 Q_3 Q_3
            + \phantom{00}\mathllap{3} Q_2 Q_2 Q_4 
        \end{array}
        \bigg){\vphantom{\Big)}}^\m{\vphantom{\Big)}}_\s \,{Dy^\s}
    \nonumber\\ {}+{}&
        (7 Q_3 Q_2 Q_2 Q_2 + 5 Q_2 Q_3 Q_2 Q_2 + 3 Q_2 Q_2 Q_3 Q_2 + Q_2 Q_2 Q_2 Q_3)^\m{}_\s \mem{dx^\s}
    \nonumber\\ {}+{}&
        (Q_2 Q_2 Q_2 Q_2)^\m{}_\s \mem{Dy^\s}
    \\
    \label{eq:Qaction-10}
    (\hhnem{({\i_ND})^8\mem \i_N R^\m{}_\n}\hhhem)\hem y^\n
    = {}\phantom{+}{}&
        (Q_{10}
            + 28 Q_8 Q_2 + 56 Q_7 Q_3 + \phantom{1}70 Q_6 Q_4 + \phantom{1}56 Q_5 Q_5 + \phantom{1}28 Q_4 Q_6 + \phantom{4}8 Q_3 Q_7 + \phantom{6}Q_2 Q_8
        )^\m{}_\s \mem{dx^\s}
    \nonumber\\ {}+{}&
        (8Q_9 \phantom{{}+28Q_8Q_2{}} 
            + 56 Q_7 Q_2 + 140 Q_6 Q_3 + 168 Q_5 Q_4 + 112 Q_4 Q_5 + 40 Q_3 Q_6 + 6 Q_2 Q_7
        )^\m{}_\s \mem{Dy^\s}
    \nonumber\\ {}+{}&
        \left(
        \begin{array}{l}
            \phantom{+}{} 
               70 Q_6 Q_2 Q_2 
            + 168 Q_5 Q_3 Q_2 
            + 168 Q_4 Q_4 Q_2
            +  80 Q_3 Q_5 Q_2
           	+  15 Q_2 Q_6 Q_2
            \\
            +  56 Q_5 Q_2 Q_3
            + 112 Q_4 Q_3 Q_3
            + \phantom{168}\llap{80} Q_3 Q_4 Q_3
            +  20 Q_2 Q_5 Q_3
            \\
            +  28 Q_4 Q_2 Q_4
            +  \phantom{168}\llap{40} Q_3 Q_3 Q_4
            +  \phantom{168}\llap{15} Q_2 Q_4 Q_4
            +   \phantom{80}\llap{8} Q_3 Q_2 Q_5
            +   \phantom{15}\llap{6} Q_2 Q_3 Q_5
            +     Q_2 Q_2 Q_6
        \end{array}
        \right){\vphantom{\Big)}}\kern-0.2em{\vphantom{\Big)}}^\m{\vphantom{\Big)}}_\s \,{dx^\s}
    \nonumber\\ {}+{}&
        \Bigg(\,\hem
        \begin{array}{l}
            \phantom{+}{} 
               56 Q_5 Q_2 Q_2 
            +  \phantom{168}\llap{112} Q_4 Q_3 Q_2
            +  \phantom{168}\llap{80} Q_3 Q_4 Q_2
            +  \phantom{80}\llap{20} Q_2 Q_5 Q_2 
            \\
            +  56 Q_4 Q_2 Q_3 
            +  \phantom{112}\llap{80} Q_3 Q_3 Q_3
            +  \phantom{168}\llap{30} Q_2 Q_4 Q_3 
            \\
            +  24 Q_3 Q_2 Q_4 
            +  \phantom{112}\llap{18} Q_2 Q_3 Q_4 
            \phantom{{}+168Q_4Q_4Q_2}
            +  \phantom{80}\llap{4} Q_2 Q_2 Q_5
        \end{array}
        \Bigg){\vphantom{\Big)}}^\m{\vphantom{\Big)}}_\s \,{Dy^\s}
    \nonumber\\ {}+{}&
        \bigg(
        \begin{array}{l}
            \phantom{+}{} 
              28 Q_4 Q_2 Q_2 Q_2 
            + 40 Q_3 Q_3 Q_2 Q_2
            + 24 Q_3 Q_2 Q_3 Q_2
            + 18 Q_2 Q_3 Q_3 Q_2
            + 15 Q_2 Q_4 Q_2 Q_2
            \\
            + \phantom{00}\llap{6} Q_2 Q_2 Q_4 Q_2 
            + \phantom{00}\llap{8} Q_3 Q_2 Q_2 Q_3
            + \phantom{00}\llap{6} Q_2 Q_3 Q_2 Q_3
	        + \phantom{00}\llap{4} Q_2 Q_2 Q_3 Q_3
		    + \phantom{00}\llap{} Q_2 Q_2 Q_2 Q_4
        \end{array}
        \bigg){\vphantom{\Big)}}^\m{\vphantom{\Big)}}_\s \,{dx^\s}
    \nonumber\\ {}+{}&
        (
	          8 Q_3 Q_2 Q_2 Q_2 
	        + 6 Q_2 Q_3 Q_2 Q_2
	        + 4 Q_2 Q_2 Q_3 Q_2
	        + 2 Q_2 Q_2 Q_2 Q_3
        )^\m{}_\s \mem{Dy^\s}
    \nonumber\\ {}+{}&
        (Q_2 Q_2 Q_2 Q_2 Q_2)^\m{}_\s \mem{dx^\s}
    \,.
\end{align}
In the attached file \texttt{Q.nb},
we have verified this result 
up to $\O(y^{10})$
by direct computations of $(\hhnem{({\i_ND})^{\ell-1}\mem \i_N R^\m{}_\n}\hhhem)\hem y^\n$
from the covariant calculus of differential forms
as in \eqref{chem-reaction}.
Eqs.~(\ref{eq:Qaction-1})-(\ref{eq:Qaction-5})
agree exactly
with Eq.\,(83) of \rcite{Vines:2014oba}.

\subsection{The Lagrangian}
\label{APP-L2}

The Lagrangian for the all-orders GDE 
in isochronous correspondence
is given in \eqref{eq:L2-tau}:
\begin{align}
	L
	\,=\,
		\frac{m}{2}\mem (Xu+Yv)^2
		- \frac{m}{2}
	\,=\,
		m\mem
		\bb{
			\frac{u^2{\mem-\,}1}{2}
			+ u\mdot v
		}
		+ \frac{1}{2}\mem mv^2
		+ m\mem \sum_{\ell=2}^\infty \frac{1}{\ell!}\mem \L_\ell
	\,.
\end{align}
In the last expression, the bracketed terms can be discarded
as a constant plus a total derivative.
The term $\frac{1}{2}\mem m v^2$, on the other hand, is the standard kinetic energy.
Hence it remains to spell out the ``interaction Lagrangian'' $\L_\ell$ at each order $\ell$,
which follows from the explicit Jacobi propagators by straightforward algebra:
\begin{align}
    \label{eq:GDEL-2}
	\L_2
	= {}\phantom{+}{}&
			u\mem Q_2 u
	\,,\\
	\label{eq:GDEL-3}
	\L_3
	= {}\phantom{+}{}&
			u\mem Q_3 u
		+
			v\mem \BB{
				4 Q_2
			}\hhem u
	\,,\\
	\label{eq:GDEL-4}
	\L_4
	= {}\phantom{+}{}&
			u\mem \BB{
				Q_4 + 4 Q_2 Q_2
			}\hem u
		+
			v\mem \BB{
				6 Q_3
			}\hhem u
		+
			v\mem \BB{
				4 Q_2
			}\hhem v
	\,,\\
	\label{eq:GDEL-5}
	\L_5
	= {}\phantom{+}{}&
			u\mem \BB{
				Q_5 + 14 Q_3 Q_2
			}\hem u
		+
			v\mem \BB{
				8 Q_4 + 16 Q_2 Q_2
			}\hhem u
		+
			v\mem \BB{
				10 Q_3
			}\hhem v
	\,,\\
	\label{eq:GDEL-6}
	\L_6
	= {}\phantom{+}{}&
			u\mem \BB{
				Q_6 + 22 Q_4 Q_2 + 14 Q_3 Q_3 + 16 Q_2 Q_2 Q_2
			}\hem u
		\nonumber\\
		{}+{}&
			v\mem \BB{
				10 Q_5 + 50 Q_3 Q_2 + 30 Q_2 Q_3
			}\hhem u
		+
			v\mem \BB{
				18 Q_4 + 16 Q_2 Q_2
			}\hhem v
	\,,\\
	\label{eq:GDEL-7}
	\L_7
	= {}\phantom{+}{}&
			u\mem \BB{
				Q_7 
				+ 32 Q_5 Q_2 + 50 Q_4 Q_3 
				+ 62 Q_3 Q_2 Q_2 + 66 Q_2 Q_3 Q_2 
			}\hem u
		\nonumber\\
		{}+{}&
			v\mem \BB{
				12 Q_6 
				+ 108 Q_4 Q_2
				+ 108 Q_3 Q_3 
				+ 52 Q_2 Q_4
				+ 64 Q_2 Q_2 Q_2
			}\hhem u
		\nonumber\\
		{}+{}&
			v\mem \BB{
				28 Q_5 + 112 Q_3 Q_2
			}\hhem v
	\,,\\
	\label{eq:GDEL-8}
	\L_8
	= {}\phantom{+}{}&
			u\mem \bb{
			\begin{aligned}[c]
				&
				Q_8
				+ 44 Q_6 Q_2 + 82 Q_5 Q_3 + 50 Q_4 Q_4
				\\
				&
				+ 114 Q_4 Q_2 Q_2 + 302 Q_3 Q_3 Q_2 + 62 Q_3 Q_2 Q_3 + 174 Q_ 2 Q_4 Q_2
				+ 64 Q_2 Q_2 Q_2 Q_2
			\end{aligned}
			}\hem u
		\nonumber\\
		{}+{}&
			v\mem \bb{
			\begin{aligned}[c]
				&
				14 Q_7 
				+ 196 Q_5 Q_2 
				+ 266 Q_4 Q_3 
				+ 210 Q_3 Q_4
				+ 84 Q_2 Q_5 
				\\
				&
				+ 238 Q_3 Q_2 Q_2
				+ 308 Q_2 Q_3 Q_2
				+ 126 Q_2 Q_2 Q_3
			\end{aligned}
			}\hhem u
		\nonumber\\
		{}+{}&
			v\mem \BB{
				40 Q_6 + 220 Q_3 Q_3 + 272 Q_4 Q_2 + 64 Q_2 Q_2 Q_2
			}\hhem v
	\,,\\
	\label{eq:GDEL-9}
	\L_9
	= {}\phantom{+}{}&
			u\mem \left(\,{
			\begin{aligned}[c]
				&
				Q_9
				+ 58 Q_7 Q_2 + 126 Q_6 Q_3 + 182 Q_5 Q_4
				\\
				&
				+ 198 Q_5 Q_2 Q_2 + 626 Q_4 Q_3 Q_2 + 238 Q_4 Q_2 Q_3 + 916 Q_3 Q_4 Q_2 + 364 Q_3 Q_3 Q_3 + 370 Q_2 Q_5 Q_2
				\\
				&
				+ 254 Q_3 Q_2 Q_2 Q_2 + 674 Q_2 Q_3 Q_2 Q_2
			\end{aligned}
			}\,\right)\hem u
		\nonumber\\
		{}+{}&
			v\mem \left(\,{
			\begin{aligned}[c]
				&
				16 Q_8 
				+ 320 Q_6 Q_2
				+ 544 Q_5 Q_3
				+ 576 Q_4 Q_4
				+ 376 Q_3 Q_5
				+ 128 Q_2 Q_6
				\\
				&
				+ 624 Q_4 Q_2 Q_2
				+ 1288 Q_3 Q_3 Q_2
				+ 488 Q_3 Q_2 Q_3
				+ 928 Q_2 Q_4 Q_2
				+ 736 Q_2 Q_3 Q_3
				+ 240 Q_2 Q_2 Q_4
				\\
				&
				+ 256 Q_2 Q_2 Q_2 Q_2
			\end{aligned}
			}\,\right)\hhem u
		\nonumber\\
		{}+{}&
			v\mem \BB{
				54 Q_7 + 552 Q_5 Q_2 + 1188 Q_4 Q_3
				+ 492 Q_3 Q_2 Q_2 + 372 Q_2 Q_3 Q_2
			}\hhem v
	\,,\\
	\label{eq:GDEL-10}
	\L_{10}
	= {}\phantom{+}{}&
			u\mem \left(\,{
			\begin{aligned}[c]
				&
				Q_{10}
				+ 74 Q_8 Q_2 + 184 Q_7 Q_3 + 308 Q_6 Q_4 + 182 Q_5 Q_5
				\\
				&
				+ 326 Q_6 Q_2 Q_2 + 1200 Q_5 Q_3 Q_2 + 436 Q_5 Q_2 Q_3 
				+ 2118 Q_4 Q_4 Q_2 + 1592 Q_4 Q_3 Q_3 + 238 Q_4 Q_2 Q_4
				\\
				&
				+ 2200 Q_3 Q_5 Q_2 + 1280 Q_3 Q_4 Q_3 + 690 Q_2 Q_6 Q_2
				+ 494 Q_4 Q_2 Q_2 Q_2 + 1540 Q_3 Q_3 Q_2 Q_2 
				\\
				&
				+ 1540 Q_3 Q_2 Q_3 Q_2 + 254 Q_3 Q_2 Q_2 Q_3 + 2226 Q_2 Q_4 Q_2 Q_2
				+ 1962 Q_2 Q_3 Q_3 Q_2
				+ 256 Q_2 Q_2 Q_2 Q_2 Q_2
			\end{aligned}
			}\,\right)\hem u
		\nonumber\\
		{}+{}&
			v\mem \left(\,{
			\begin{aligned}[c]
				&
				18 Q_9 
				+ 486 Q_7 Q_2
				+ 990 Q_6 Q_3
				+ 1302 Q_5 Q_4
				+ 1134 Q_4 Q_5
				+ 630 Q_3 Q_6
				+ 186 Q_2 Q_7
				+ 1374 Q_5 Q_2 Q_2 
				\\
				&
				+ 3726 Q_4 Q_3 Q_2
				+ 1350 Q_4 Q_2 Q_3
				+ 4320 Q_3 Q_4 Q_2
				+ 3240 Q_3 Q_3 Q_3
				+ 966 Q_3 Q_2 Q_4
				+ 2220 Q_2 Q_5 Q_2
				\\
				&
				+ 2580 Q_2 Q_4 Q_3
				+ 1602 Q_2 Q_3 Q_4
				+ 438 Q_2 Q_2 Q_5
				+ 1002 Q_3 Q_2 Q_2 Q_2
				+ 1674 Q_2 Q_3 Q_2 Q_2
				\\
				&
				+ 1422 Q_2 Q_2 Q_3 Q_2
				+ 510 Q_2 Q_2 Q_2 Q_3
			\end{aligned}
			}\,\right)\hhem u
		\nonumber\\
		{}+{}&
			v\mem \bb{
			\begin{aligned}[c]
				&
				70 Q_8 + 1000 Q_6 Q_2 + 2660 Q_5 Q_3 + 1764 Q_4 Q_4
				+ 1356 Q_4 Q_2 Q_2 + 3260 Q_3 Q_3 Q_2 + 980 Q_3 Q_2 Q_3 
				\\
				&
				+ 1300 Q_2 Q_4 Q_2
				+ 256 Q_2 Q_2 Q_2 Q_2
			\end{aligned}
			}\hhem v
	\,.
\end{align}
Eqs.~(\ref{eq:GDEL-2})-(\ref{eq:GDEL-5})
agree flawlessly with Eq.\,(5) in Vines \cite{Vines:2014oba}.

\pagebreak

\subsection{The GDE}
\label{APP-2GDE}
The GDE, in the second-order formulation, is given in \eqref{2GDE}.
The definition of $X^\m{}_\s$ and $Y^\m{}_\s$
is given in \eqref{sol}.
The definition of $\acute{X}^\m{}_{\n\r}$ and $\acute{Y}^\m{}_{\n\r}$
is given in \eqref{sol-acute}.
For the reader's sake, we explicitly spell them out at low orders:\nem
\begin{subequations}
\label{AllXYs5}
\begin{align}
	\label{AllX5}
	X^\m{}_\s
	\mem&=\mem
		\delta^\m{}_\s
		+ \frac{1}{2!}\mem (Q_2)^\m{}_\s
		+ \frac{1}{3!}\mem (Q_3)^\m{}_\s
		+ \frac{1}{4!}\mem (Q_4 + Q_2 Q_2)^\m{}_\s
		+ \frac{1}{5!}\mem (Q_5 + 3 Q_3 Q_2 + Q_2 Q_3)^\m{}_\s
		+ \O(y^6)
	\,,\\
	\label{AllY5}
	\phantom{\acute{X}^\m{}_{\n\s}{}}
	\mathllap{Y^\m{}_\s}
	\mem&=\mem
		\delta^\m{}_\s
		+ \frac{1}{3!}\mem (Q_2)^\m{}_\s
		+ \frac{1}{4!}\mem (2 Q_3)^\m{}_\s
		+ \frac{1}{5!}\mem (3 Q_4 + Q_2 Q_2)^\m{}_\s
		+ \frac{1}{6!}\mem (4 Q_5 + 4 Q_3 Q_2 + 2 Q_2 Q_3)^\m{}_\s
		+ \O(y^6)
	\,,
\end{align}
\end{subequations}
\vspace{-1.6\baselineskip}
\begin{subequations}
\begin{align}
	\label{AllacX5}
	\acute{X}^\m{}_{\n\s}
	\mem&=\mem
	\mathrlap{
	\begin{aligned}[t]
		  \frac{1}{1!}\mem (\acute{Q}_2)^\m{}_{\n\s}
		&+ \frac{1}{2!}\mem (\acute{Q}_3)^\m{}_{\n\s}
		+ \frac{1}{3!}\mem \BB{
			(\acute{Q}_4)^\m{}_{\n\s} + (\acute{Q}_2)^\m{}_{\n\l}\mem (Q_2)^\l{}_\s
		}\\
		&+ \frac{1}{4!}\mem \BB{
			(\acute{Q}_5)^\m{}_{\n\s} 
			+ 3 (\acute{Q}_3)^\m{}_{\n\l}\mem (Q_2)^\l{}_\s
			+ (\acute{Q}_2)^\m{}_{\n\l}\mem (Q_3)^\l{}_\s
		}
		+ \O(y^5)
		\,,
	\end{aligned}
	}
	\phantom{{}
			\delta^\m{}_\s
			+ \frac{1}{3!}\mem (Q_2)^\m{}_\s
			+ \frac{1}{4!}\mem (2 Q_3)^\m{}_\s
			+ \frac{1}{5!}\mem (3 Q_4 + Q_2 Q_2)^\m{}_\s
			+ \frac{1}{6!}\mem (4 Q_5 + 4 Q_3 Q_2 + 2 Q_2 Q_3)^\m{}_\s
			+ \O(y^6)
		\,,
	}
	\\
	\label{AllacY5}
	\acute{Y}^\m{}_{\n\s}
	\mem&=\mem
	\begin{aligned}[t]
		  \frac{1}{2!}\mem (\acute{Q}_2)^\m{}_{\n\s}
		&+ \frac{1}{3!}\mem (2 \acute{Q}_3)^\m{}_{\n\s}
		+ \frac{1}{4!}\mem \BB{
			3 (\acute{Q}_4)^\m{}_{\n\s} + (\acute{Q}_2)^\m{}_{\n\l}\mem (Q_2)^\l{}_\s
		}\\
		&+ \frac{1}{5!}\mem \BB{
			4 (\acute{Q}_5)^\m{}_{\n\s} 
			+ 4 (\acute{Q}_3)^\m{}_{\n\l}\mem (Q_2)^\l{}_\s
			+ 2 (\acute{Q}_2)^\m{}_{\n\l}\mem (Q_3)^\l{}_\s
		}
		+ \O(y^5)
		\,.
	\end{aligned}
\end{align}
\end{subequations}

Next, we need to compute 
the inverse $(Y^{-1})^\m{}_\n$ of the Jacobi propagator $Y^\m{}_\n$
in \eqref{AllY5},
which is viable by geometric series expansion around $\delta^\m{}_\n$:
\begin{align}
	Y^{-1}
	\mem=\mem
		\id
		&
		- \frac{1}{3!}\mem \BB{
			Q_2
		}
		- \frac{1}{4!}\mem \BB{
			2\mem Q_3
		}
		- \frac{1}{5!}\mem \bb{
			3\mem Q_4 - \frac{7}{3}\mem Q_2 Q_2
		}
		- \frac{1}{6!}\mem \bb{
			4\mem Q_5 - 8\mem Q_2 Q_3 - 6\mem Q_3 Q_2
		}
		\\\nonumber
		&
		- \frac{1}{7!}\mem \bb{
			5\mem Q_6 
			- 11\mem Q_4 Q_2 - 25\mem Q_3 Q_3 - 18\mem Q_2 Q_4 
			+ \frac{31}{3}\mem Q_2 Q_2 Q_2
		}
		\\\nonumber
		&
		- \frac{1}{8!}\mem \bb{
			6\mem Q_7 
			- \frac{52}{3}\mem Q_5 Q_2
			- 54\mem Q_4 Q_3 
			- 66\mem Q_3 Q_4
			- \frac{100}{3}\mem Q_2 Q_5
			+ 34\mem Q_3 Q_2 Q_2
			+ \frac{124}{3}\mem Q_2 Q_3 Q_2 
			+ \frac{146}{3}\mem Q_2 Q_2 Q_3 
		}
		\\\nonumber
		&
		- \frac{1}{9!}\mem \lrp{
		\begin{aligned}[c]
			&
			7\mem Q_8 
			- 25\mem Q_6 Q_2
			- 98\mem Q_5 Q_3 
			- \frac{819}{5}\mem Q_4 Q_4 
			- 140\mem Q_3 Q_5 
			- 55\mem Q_2 Q_6 
			+ \frac{387}{5}\mem Q_4 Q_2 Q_2
			+ 160\mem Q_3 Q_3 Q_2
			\\
			&
			+ 182\mem Q_3 Q_2 Q_3
			+ 106\mem Q_2 Q_4 Q_2
			+ 226\mem Q_2 Q_3 Q_3
			+ \frac{717}{5}\mem Q_2 Q_2 Q_4
			- \frac{381}{5}\mem Q_2 Q_2 Q_2 Q_2
		\end{aligned}
		}
		\\\nonumber
		&
		- \frac{1}{10!}\mem \lrp{
		\begin{aligned}[c]
			&
			8\mem Q_9 
			- 34\mem Q_7 Q_2
			- 160\mem Q_6 Q_3 
			- 336\mem Q_5 Q_4 
			- 392\mem Q_4 Q_5 
			- 260\mem Q_3 Q_6 
			- 84\mem Q_2 Q_7 
			\\
			& 
			+ 148\mem Q_5 Q_2 Q_2 
			+ 418\mem Q_4 Q_3 Q_2 + 464\mem Q_4 Q_2 Q_3  
			+ 470\mem Q_3 Q_4 Q_2 + 980\mem Q_3 Q_3 Q_3 + 600\mem Q_3 Q_2 Q_4
			\\
			&
			+ 220\mem Q_2 Q_5 Q_2 + 660\mem Q_2 Q_4 Q_3 + 756\mem Q_2 Q_3 Q_4 + 336\mem Q_2 Q_2 Q_5 
			\\
			&
			- 310\mem Q_3 Q_2 Q_2 Q_2 - 368\mem Q_2 Q_3 Q_2 Q_2 - 394\mem Q_2 Q_2 Q_3 Q_2 - 452\mem Q_2 Q_2 Q_2 Q_3 
		\end{aligned}
		}
		\\\nonumber
		&
		- \frac{1}{11!}\mem \lrp{
		\begin{aligned}[c]
			&
			9\mem Q_{10} 
			- \frac{133}{3}\mem Q_8 Q_2
			- 243\mem Q_7 Q_3
			- 612\mem Q_6 Q_4
			- 896\mem Q_5 Q_5
			- 810\mem Q_4 Q_6
			- 441\mem Q_3 Q_7
			- \frac{364}{3}\mem Q_2 Q_8
			\\
			& 
			+ \frac{763}{3}\mem Q_6 Q_2 Q_2
			+ \frac{2702}{3}\mem Q_5 Q_3 Q_2
			+ 982\mem Q_5 Q_2 Q_3
			+ 1383\mem Q_4 Q_4 Q_2 
			+ 2835\mem Q_4 Q_3 Q_3
			\\
			& 
			+ 1692\mem Q_4 Q_2 Q_4
			+ \frac{3290}{3}\mem Q_3 Q_5 Q_2
			+ 3240\mem Q_3 Q_4 Q_3
			+ 3627\mem Q_3 Q_3 Q_4
			+ 1554\mem Q_3 Q_2 Q_5
			\\
			&
			+ \frac{1205}{3}\mem Q_2 Q_6 Q_2
			+ \frac{4610}{3}\mem Q_2 Q_5 Q_3
			+ 2472\mem Q_2 Q_4 Q_4
			+ \frac{5936}{3}\mem Q_2 Q_3 Q_5
			+ \frac{2050}{3}\mem Q_2 Q_2 Q_6
			\\
			&
			- 855\mem Q_4 Q_2 Q_2 Q_2
			- \frac{5173}{3}\mem Q_3 Q_3 Q_2 Q_2
			- \frac{5348}{3}\mem Q_3 Q_2 Q_3 Q_2
			- 2028\mem Q_3 Q_2 Q_2 Q_3
			\\
			& 
			- \frac{3358}{3}\mem Q_2 Q_4 Q_2 Q_2
			- \frac{6494}{3}\mem Q_2 Q_3 Q_3 Q_2
			- \frac{7262}{3}\mem Q_2 Q_3 Q_2 Q_3
			- \frac{3787}{3}\mem Q_2 Q_2 Q_4 Q_2
			\\
			&
			- \frac{7945}{3}\mem Q_2 Q_2 Q_3 Q_3
			- 1636\mem Q_2 Q_2 Q_2 Q_4	
			+ \frac{2555}{3}\mem Q_2 Q_2 Q_2 Q_2 Q_2
		\end{aligned}
		}
	\,.
\end{align}

Finally,
the right-hand side of \eqref{2GDE}
should be computed.
The algebraic terms, 
\smash{$\acX^\m{}_{\n\r}\mem X^\n{}_\s$},
\smash{$\acX^\m{}_{\n\r}\mem Y^\n{}_\s$},
\smash{$\acY^\m{}_{\n\r}\mem X^\n{}_\s$},
and
\smash{$\acY^\m{}_{\n\r}\mem Y^\n{}_\s$},
are straightforward to evaluate
from \eqref{AllXYs5}.
The evaluation of the differential terms,
however,
is relatively less trivial.
The issue in particular is the evaluation of the $y$-derivative.
When
a $y$-derivative hits a $Q$-tensor,
it inserts $v$ in all possible positions in the string of $y$-vectors.
Such terms can be gathered and simplified
by commuting covariant derivatives
and using Bianchi identities of the Riemann tensor
in various ways.

For the part linear in both $u$ and $v$,
this issue is simply avoided 
by virtue ot
the zero-torsion identity in \eqref{torid-b}.
Hence it suffices to massage
the part quadratic in $v$,
which is the third term in the right-hand side of \eqref{2GDE}.

As the result of this simplification,
we obtain the final expression
as a sum of terms
that strictly conform to the following particular tensor structures:
\begin{subequations}
\label{structures}
\begin{align}
	\label{structure-D}
	\text{Derivative}:\quad
	&
	(Q_{\c_1} \cdots Q_{\c_r})^\m{}_\n
	\,
	(Q_\ell)^\n{}_{\k|\r}
	\,
	(Q_{\a_1} \cdots Q_{\a_p})^\k{}_\s
	\,,\\
	\label{structure-B}
	\text{Branching}:\quad
	&
	(Q_{\c_1} \cdots Q_{\c_r})^\m{}_\n
	\,
	(\acQ_\ell)^\n{}_{\l\k}
	\,
	(Q_{\a_1} \cdots Q_{\a_p})^\k{}_\s
	\,
	(Q_{\b_1} \cdots Q_{\b_q})^\l{}_\r
	\,.
\end{align}
\end{subequations}
Here, we have denoted
\begin{align}
	\label{DQ-def}
	(Q_\ell)^\m{}_{\s|\r}
	\,:=\,
		y^{\k_1}{\cdots}y^{\k_{\ell-1}}\mem
	    R^\m{}_{\k_1\k_2\s;\k_3;\cdots;\k_{\ell-1};\r}(x)
	\quad\text{for}\quad
	\ell \geq 3
	\,.
\end{align}
Crucially, 
the free index $\r$ in \eqref{DQ-def},
which can be contracted with $u$ or $v$, for instance,
is made to
describe the last (outmost) covariant derivative
acting on the Riemann tensor.
This is always possible by commuting covariant derivatives,
while
employing the Bianchi identities
facilitates
rewriting the remaining terms 
strictly in terms of the $Q$- and $\acQ$-tensors.

The final results are contained in the attached Mathematica notebook \texttt{All.nb}.
For the output data type, we have chosen to represent the tensor structures in \eqrefs{structure-D}{structure-B}
as lists consisting of integers and symbols $\i$, $\D$.
For example,
\begin{subequations}
\label{seq}
\begin{align}
	\label{seq-derivative}
	(Q_3 Q_2)^\m{}_\n\mem 
	(Q_3)^\n{}_{\s|\r}\mem
	u^\r\mem
	(Q_5Q_4Q_4Q_6v)^{\s}
	\quad&\rightsquigarrow\quad
		\texttt{% 
			\string{% 
				3,2,$\D$[u][3],5,4,4,6,v% 
			\string}% 
		}
	\,,\\
	\label{seq-branching}
	(Q_3 Q_2)^\m{}_\n\mem 
	(\acQ_3)^\n{}_{\r\s}\mem
	(Q_2Q_3Q_3v)^{\r}\mem
	(Q_5Q_4Q_4Q_6u)^{\s}
	\quad&\rightsquigarrow\quad
		\texttt{% 
			\string{% 
				3,2,$\i$[2,3,3,v][3],5,4,4,6,u% 
			\string}% 
		}
	\,.
\end{align}
\end{subequations}

We have also implemented visual output,
which might be easier to process for some.
Recalling our earlier exploration of Penrose graphical notation in Appendix~\ref{organic-chemistry},
it is natural to represent the tensor structures in Eqs.\,(\ref{structure-D}) and (\ref{structure-B}) as
``carbon chains'':
\begin{subequations}
\label{graphs}
\begin{align}
	\label{graph-derivative}
	(Q_3 Q_2)^\m{}_\n\mem 
	(Q_3)^\n{}_{\s|\r}\mem
	u^\r\mem
	(Q_5Q_4Q_4Q_6v)^{\s}
	\quad&\rightsquigarrow\quad
	\adjustbox{valign=t,raise=6pt}{\begin{tikzpicture}
		\coordinate (o) at (0,0);
		\coordinate (d) at (-30:0.4);
		\coordinate (u) at ( 30:0.4);
		\coordinate (k) at (0,-0.45);
		\coordinate (C0) at (o);
		\coordinate (C1) at ($(C0)+(d)$);
		\coordinate (C2) at ($(C1)+(u)$);
		\coordinate (C3) at ($(C2)+(d)$);
		\coordinate (C4) at ($(C3)+(u)$);
		\coordinate (C5) at ($(C4)+(d)$);
		\coordinate (C6) at ($(C5)+(u)$);
		\coordinate (C7) at ($(C6)+(d)$);
		\coordinate (C8) at ($(C7)+(u)$);
		\coordinate (K0) at ($(C3)+(k)$);
		\draw[line] ($(C3)+(0.033,0)$)--($(K0)+(0.033,0)$);
		\draw[line] ($(C3)-(0.033,0)$)--($(K0)-(0.033,0)$);
		\draw[line] (C0)--(C1)--(C2)--(C3)--(C4)--(C5)--(C6)--(C7)--(C8);
		\node[L] at (C0) {$\m$};
		\node[D] at (C1) {3};
		\node[U] at (C2) {2};
		\node[D] at (C3) {\quad\mem 3};
		\node[U] at (C4) {5};
		\node[D] at (C5) {4};
		\node[U] at (C6) {4};
		\node[D] at (C7) {6};
		\node[Y] at (C8) {};
		\node[X] at (K0) {};
	\end{tikzpicture}}
	\,\,\,
	\,,\\
	\label{graph-branching}
	(Q_3 Q_2)^\m{}_\n\mem 
	(\acQ_3)^\n{}_{\r\s}\mem
	(Q_2Q_3Q_3v)^{\r}\mem
	(Q_5Q_4Q_4Q_6u)^{\s}
	\quad&\rightsquigarrow\quad
	\adjustbox{valign=t,raise=6pt}{\begin{tikzpicture}
		\coordinate (o) at (0,0);
		\coordinate (d) at (-30:0.4);
		\coordinate (u) at ( 30:0.4);
		\coordinate (b) at (0,-0.7);
		\coordinate (C0) at (o);
		\coordinate (C1) at ($(C0)+(d)$);
		\coordinate (C2) at ($(C1)+(u)$);
		\coordinate (C3) at ($(C2)+(d)$);
		\coordinate (C4) at ($(C3)+(u)$);
		\coordinate (C5) at ($(C4)+(d)$);
		\coordinate (C6) at ($(C5)+(u)$);
		\coordinate (C7) at ($(C6)+(d)$);
		\coordinate (C8) at ($(C7)+(u)$);
		\coordinate (B0) at ($(C3)+(b)$);
		\coordinate (B1) at ($(B0)+(d)$);
		\coordinate (B2) at ($(B1)+(u)$);
		\coordinate (B3) at ($(B2)+(d)$);
		\coordinate (B4) at ($(B3)+(u)$);
		\draw[bine] (C3)--(B0);
		\draw[line] (C0)--(C1)--(C2)--(C3)--(C4)--(C5)--(C6)--(C7)--(C8);
		\draw[line] (B0)--(B1)--(B2)--(B3)--(B4);
		\node[L] at (C0) {$\m$};
		\node[D] at (C1) {3};
		\node[U] at (C2) {2};
		\node[D] at (C3) {3};
		\node[U] at (C4) {5};
		\node[D] at (C5) {4};
		\node[U] at (C6) {4};
		\node[D] at (C7) {6};
		\node[X] at (C8) {};
		\node[ ] at (B0) {};
		\node[D] at (B1) {2};
		\node[U] at (B2) {3};
		\node[D] at (B3) {3};
		\node[Y] at (B4) {};
	\end{tikzpicture}}
	\,\,\,
	\,.
\end{align}
\end{subequations}
Here, the vectors $u$ and $v$ are represented as white and blue blobs,
respectively:
\begin{align}
	u^\m
	\quad\rightsquigarrow\quad
	\adjustbox{valign=c}{\begin{tikzpicture}
		\coordinate (o) at (0,0);
		\coordinate (C0) at (o);
		\coordinate (C1) at ($(C0)+(0.6,0)$);
		\draw[line] (C0)--(C1);
		\node[L] at (C0) {$\m$};
		\node[X] at (C1) {};
	\end{tikzpicture}}
	\,\,\,\,,\qquad
	v^\m
	\quad\rightsquigarrow\quad
	\adjustbox{valign=c}{\begin{tikzpicture}
		\coordinate (o) at (0,0);
		\coordinate (C0) at (o);
		\coordinate (C1) at ($(C0)+(0.6,0)$);
		\draw[line] (C0)--(C1);
		\node[L] at (C0) {$\m$};
		\node[Y] at (C1) {};
	\end{tikzpicture}}
	\,\,\,\,.
\end{align}
And of course, 
we have denoted matrix products of $Q$-tensors as
\begin{align}
	(Q_3Q_2Q_3Q_5)^\m{}_\n
	\,=\,
		(Q_3)^\m{}_{\l_1}\mem (Q_2)^{\l_1}{}_{\l_2}\mem (Q_3)^{\l_2}{}_{\l_3}\mem (Q_5)^{\l_3}{}_\n
	\quad&\rightsquigarrow\quad
		\adjustbox{valign=c}{\begin{tikzpicture}
			\coordinate (o) at (0,0);
			\coordinate (d) at (-30:0.4);
			\coordinate (u) at ( 30:0.4);
			\coordinate (C0) at (o);
			\coordinate (C1) at ($(C0)+(d)$);
			\coordinate (C2) at ($(C1)+(u)$);
			\coordinate (C3) at ($(C2)+(d)$);
			\coordinate (C4) at ($(C3)+(u)$);
			\coordinate (C5) at ($(C4)+(d)$);
			\draw[line] (C0)--(C1)--(C2)--(C3)--(C4)--(C5);
			\node[L] at (C0) {$\m$};
			\node[D] at (C1) {3};
			\node[U] at (C2) {2};
			\node[D] at (C3) {3};
			\node[U] at (C4) {5};
			\node[R] at (C5) {$\n$};
		\end{tikzpicture}}
	\,.
\end{align}
Amusingly,
these visualizations can be conveniently implemented
with the built-in \texttt{MoleculePlot3D} function
in Mathematica.
For instance, 
in the attached file \texttt{All.nb},
we have denoted
\begin{subequations}
\label{mols}
\begin{align}
\label{mol-D}
	(Q_3)^\m{}_\n\mem (Q_4)^\n{}_{\s|\r}\mem (Q_2 Q_2 Q_5 v)^\s
	\, u^\r
	\quad&\rightsquigarrow\quad
	\includegraphics[scale=0.5,valign=c]{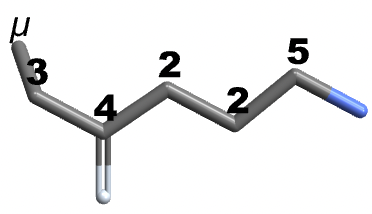}
	\,\,\,
	\,,\\
\label{mol-B}
	(Q_3)^\m{}_\n\mem (\acQ_3)^\n{}_{\r\s}\mem (Q_2 v)^\r
	\, (Q_5 u)^\s
	\quad&\rightsquigarrow\quad
	\includegraphics[scale=0.5,valign=c]{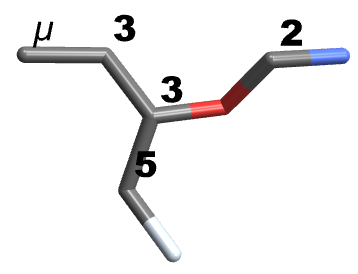}
	\,\,\,
	\,.
\end{align}
\end{subequations}
In sum, 
a soup of molecules is produced from
a sequence of tensor manipulations,
encoding the laws of gravitational dynamics for a free-falling particle in space.
Each of the molecules
exhibits the characteristic triple-strand structure 
shown in Eqs.\,(\ref{structures}), (\ref{seq}), (\ref{graphs}), or (\ref{mols}),
describing
carbon chains built out of Riemann tensors.
While this chemical language can
appear as
a case of avant-garde art,
we also realize that
it could serve as an
intuitive and efficient notation
specialized for the tensors relevant to the GDE, in fact.

\newpage

Finally,
below, we enumerate the GDE (as the right-hand side of $-Dy^\m\nem/d\t = \cdots$) at each order
in the chemical notation,
up to $\O(y^6)$.

\begin{align}
	&\includegraphics[scale=0.52,valign=t]{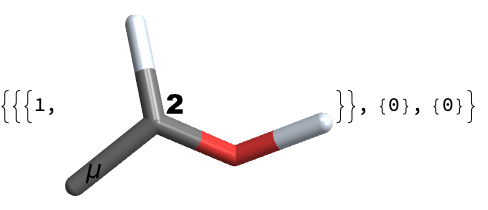}
\\[1.52\baselineskip]
	&\includegraphics[scale=0.52,valign=t]{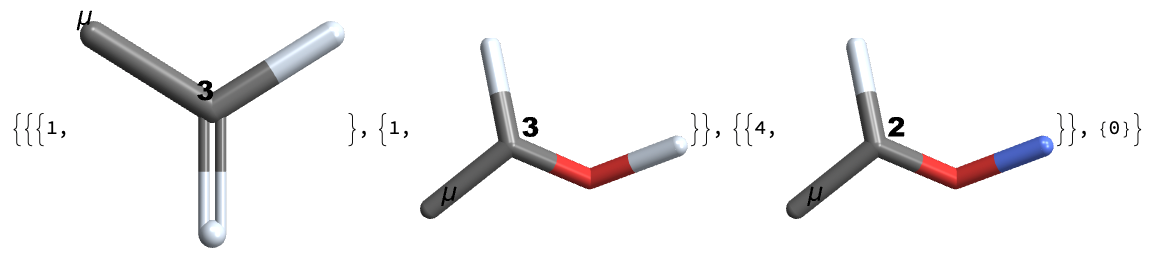}
\\[1.52\baselineskip]
	&\includegraphics[scale=0.52,valign=t]{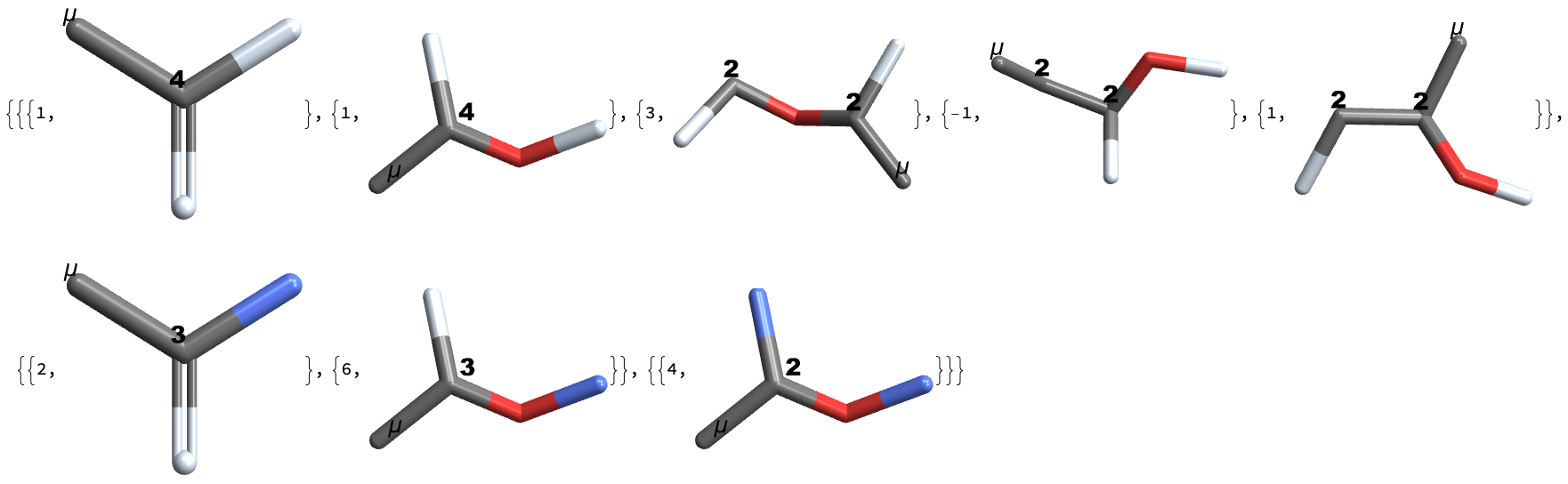}
\\[1.52\baselineskip]
	&\includegraphics[scale=0.52,valign=t]{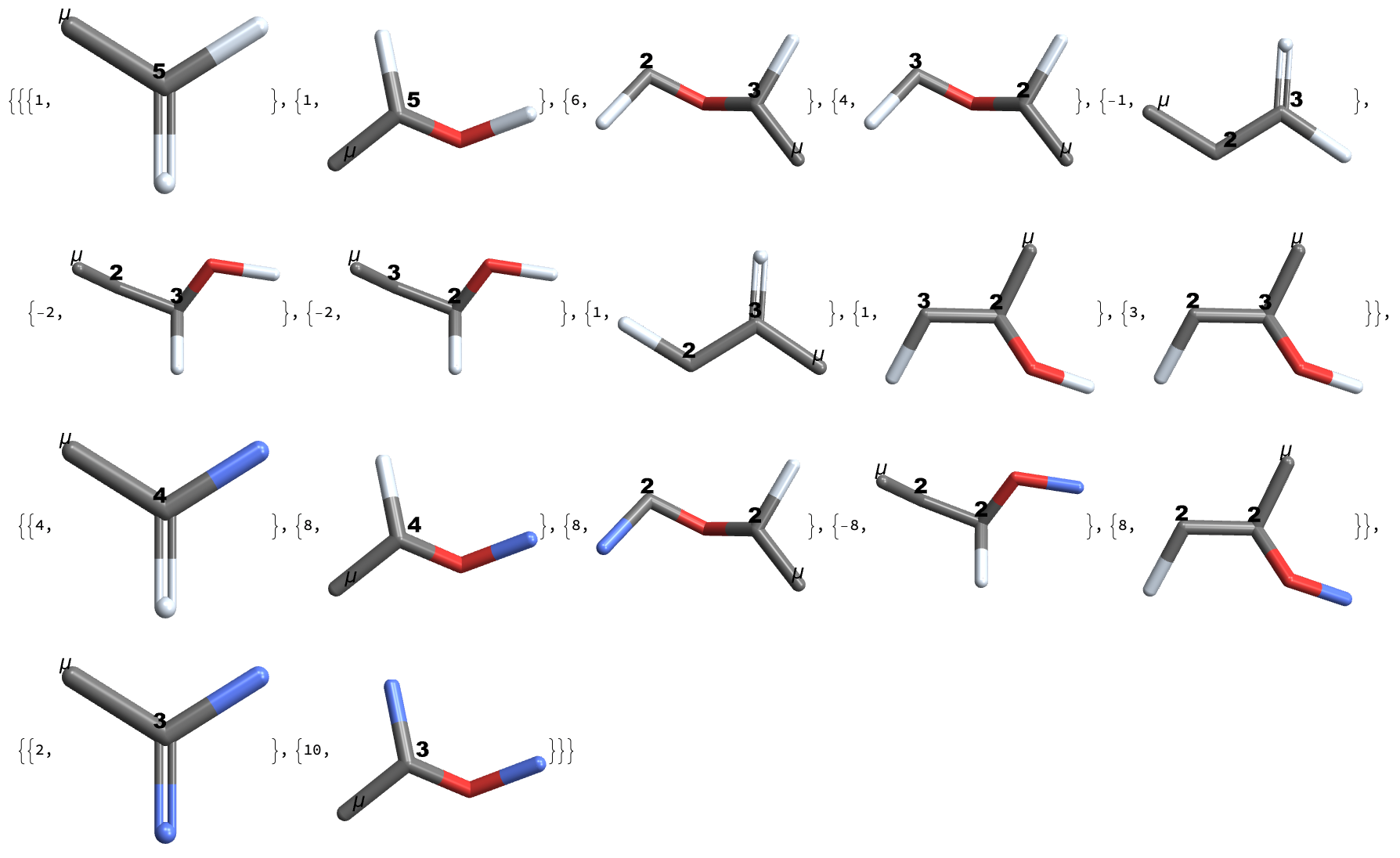}
\end{align}

\newpage

\begin{subequations}
\begin{align}
	&\includegraphics[scale=0.3,valign=t]{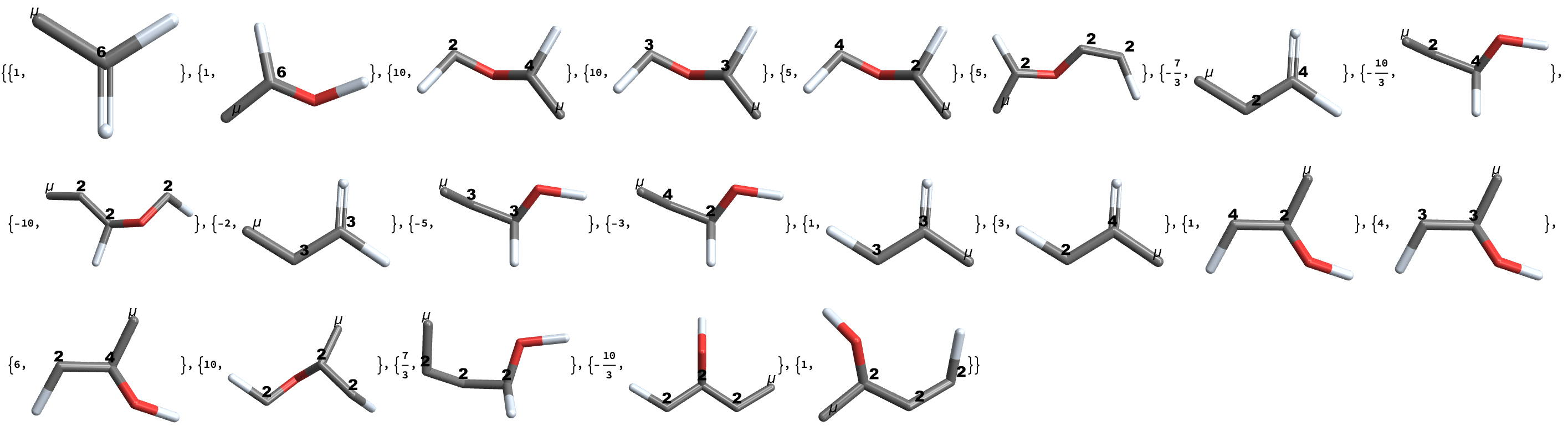}
\\[0.15\baselineskip]
	&\includegraphics[scale=0.3,valign=t]{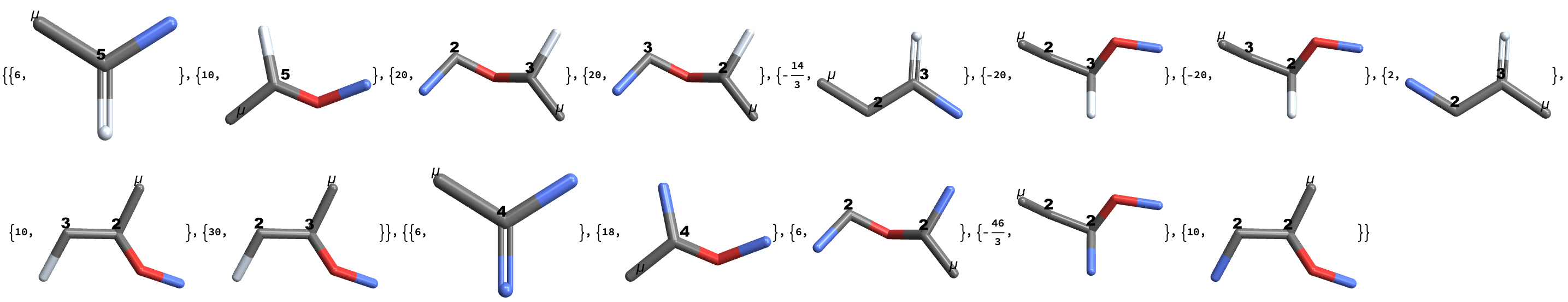}
\end{align}
\end{subequations}

\begin{subequations}
\begin{align}
	&\includegraphics[scale=0.3,valign=t]{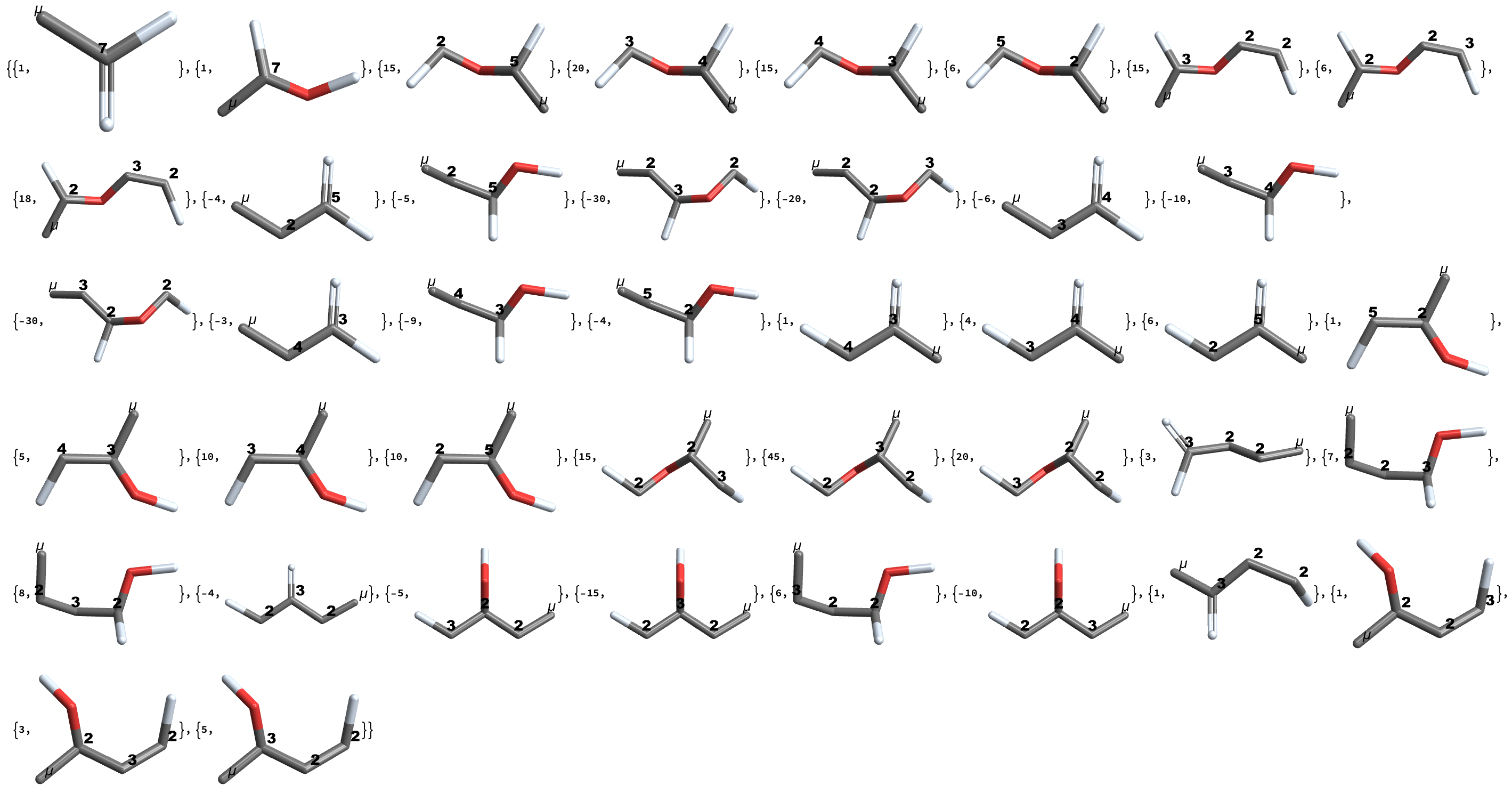}
\\[0.15\baselineskip]
	&\includegraphics[scale=0.3,valign=t]{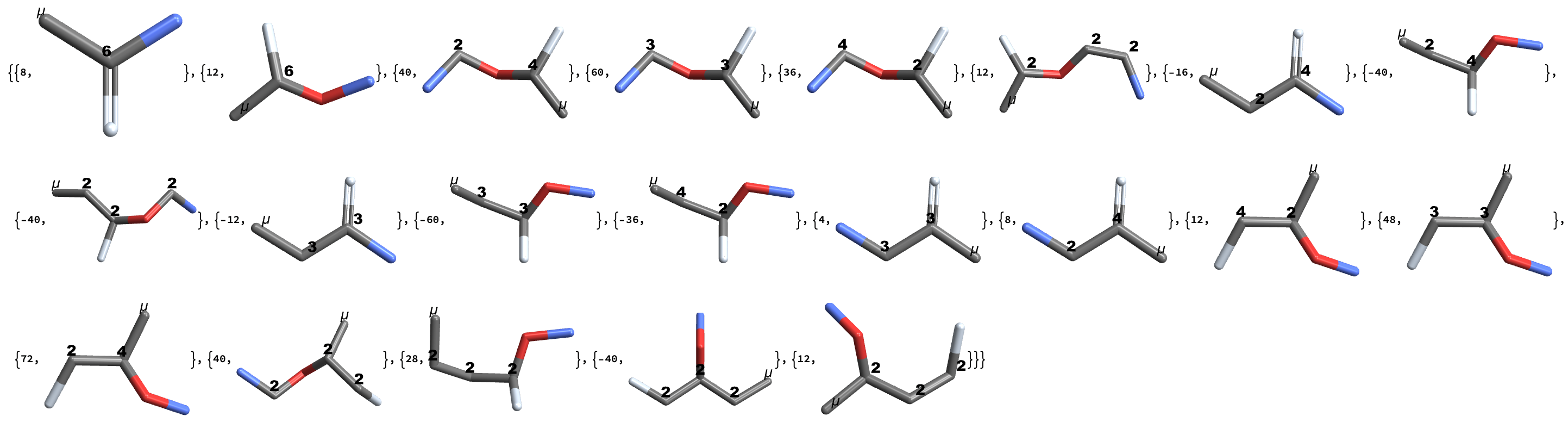}
\\[0.15\baselineskip]
	&\includegraphics[scale=0.3,valign=t]{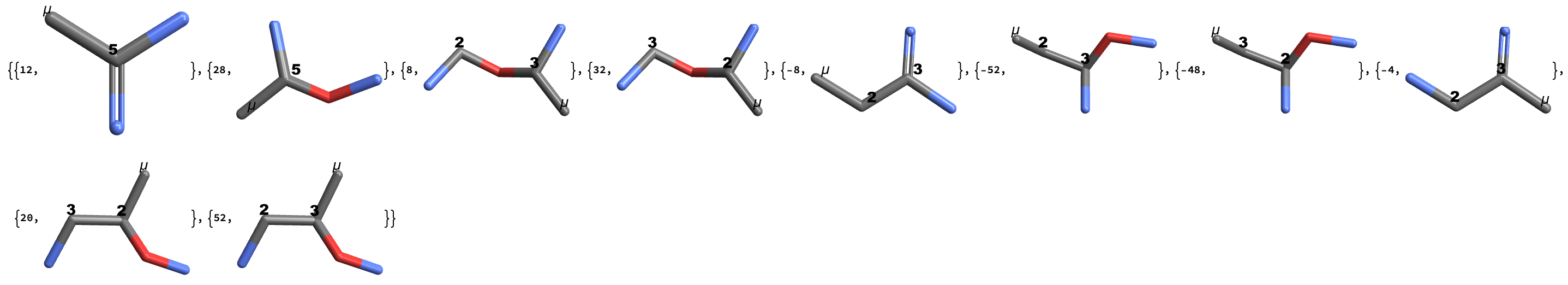}
\end{align}
\end{subequations}

\newpage

\section{Gauge-Covariant Translations in Nonabelian Gauge Theory}
\label{gaugetheory}

It should be remarked that
our framework applies to 
not only gravity 
but also
nonabelian gauge theories.
Suppose a nonabelian gauge theory in a $d$-dimensional flat spacetime $\mflat$
with gauge group $G$.
Let $A^i{}_j = A^i{}_{j\m}(x)\mem dx^\m$ be the gauge connection
and let $F^i{}_j = dA^i{}_j + A^i{}_k \swedge A^k{}_j$
be its curvature,
where $i,j,\cdots = 1,2,\cdots,N$ are the fundamental indices.
For example, suppose a vector bundle $E$ over $\mflat$
whose typical fiber is $\mathbb{C}^N$.
This bundle is coordinatized by
$x^\m$ and $\psi^i$,
associated respectively with the base and fiber.

Our tangent bundle formalism
then considers the larger bundle
$\P = T\mflat \oplus E$,
which is locally isomorphic to
$\mathbb{R}^{2d} \times \mathbb{C}^N$.
Local trivialization equips $\P$ with
coordinates $x^\m$, $y^\m$, and $\psi^i$.
In $\P$, the generator of gauge-covariant translations 
is given by the vector field
\begin{align}
	N
	\,=\,
		y^\r\,
		\bb{
			\frac{\partial}{\partial x^\r}
			- A^i{}_{j\r}(x)\mem \psi^j
			\,\frac{\partial}{\partial \psi^i}
		}
	\,.
\end{align}
Again, this describes a horizontal vector field
due to the Ehresmann \cite{ehresmann1948connexions}
notion of the connection $A$.
Recalling \eqref{eq:diffeq-XY},
one derives the set of first-order differential equations
it encodes,
which leads to the conclusion that
it generates gauge-covariant translations:
$x^\m \mapsto x^\m + y^\m$,
$y^\m \mapsto y^\m$,
and
$\psi^i \mapsto \psi^{i'} = W^{i\sprime}{}_i\mem \psi^i$.
Here, the Wilson line $W^{i\sprime}{}_i$ is given by the path-ordered exponential,
\begin{align}
	\label{wilsonline}
	\mathrm{P} \exp \bb{\nem
		{-\nem \int_0^1 ds\,\,
			A_\r(x + ys)\,
				y^\r
		}
	}
	\,,
\end{align}
which describes the parallel transport
along the straight path
from $x$ to $x+y$.
Under gauge transformations, \eqref{wilsonline} transforms bilocally as
$W^{i\sprime}{}_i \mapsto \Lambda^{i\sprime}{}_{j'}(z)\, W^{j\sprime}{}_j\, (\Lambda^{-1}\hnem(x)\hhnem)^j{}_i$.

Now consider the ``$\pounds_N^D$ sequence''
of the one-form $D\psi^i = d\psi^i + A^i{}_{j\r}(x)\mem \psi^j\mem dx^\r$.
\begin{align}
	{\begin{tikzpicture}
			\node[empty] (O) at (0,0) {};
			\node[empty] (X) at (3.0, 0) {};
			\node[empty] (Y) at (0, -0.9) {};
			\node[w] (a00) at ($(O)$) {$D\psi^i$};
			\node[w] (a01) at ($(O)+1*(X)$) {$0$};
			\node[w] (a10) at ($(O)+1*(Y)$) {$\i_N F^i{}_j\mem \psi^j$};
			\node[w] (a11) at ($(O)+1*(Y)+1*(X)$) {$0$};
			\node[w] (a20) at ($(O)+2*(Y)$) {$\i_N D \i_N F^i{}_j\mem \psi^j$};
			\node[w] (a21) at ($(O)+2*(Y)+1*(X)$) {$0$};
			\node[w] (a30) at ($(O)+3*(Y)$) {$\vdots$};
			\node[w] (phantom-a00) at ($(O)$) {};
			\node[w] (phantom-a01) at ($(O)+1*(X)$) {};
			\node[w] (phantom-a10) at ($(O)+1*(Y)$) {};
			\node[w] (phantom-a11) at ($(O)+1*(Y)+1*(X)$) {};
			\node[w] (phantom-a20) at ($(O)+2*(Y)$) {};
			\node[w] (phantom-a21) at ($(O)+2*(Y)+1*(X)$) {};
			\node[w] (phantom-a30) at ($(O)+3*(Y)$) {};
			\draw[->] (a00)--(a01) node[midway,above] {\scriptsize $D\mem\i_N$};
			\draw[->] (a10)--(a11) node[] {};
			\draw[->] (a20)--(a21) node[] {};
			\draw[->] (phantom-a00)--(phantom-a10) node[midway,left] {\scriptsize $\i_ND$};
			\draw[->] (phantom-a10)--(phantom-a20) node[] {};
			\draw[->] (phantom-a20)--(phantom-a30) node[] {};
	\end{tikzpicture}}
\end{align}
This derives 
\begin{subequations}
\label{Dpsi}
\begin{align}
	\label{Dpsi-series}
	\mathe^{\pounds_N^D} D\psi^i
	\,&=\,
		D\psi^i
		+ \sum_{\ell=1}^\infty\,
			\frac{1}{\ell!}\,
			(\i_ND)^{\ell-1} \i_N F^i{}_j\mem \psi^j
	\,,\\
	\label{Dpsi-series-expanded}
	\,&=\,
		D\psi^i
		+ \sum_{\ell=1}^\infty\,
			\frac{1}{\ell!}\,
			\BB{
				(P_\ell)^i{}_{j\s}\mem dx^\s
				+ (\ell {\,-\mem} 1)\mem (P_{\ell-1})^i{}_{j\s}\mem dy^\s
			}
			\mem \psi^j
	\,,
\end{align}
\end{subequations}
where the ``$P$-tensors'' are defined for $\ell \geq 1$ as
\begin{align}
    \label{eq:def-Pten}
    (P_\ell)^i{}_{j\s}
    {}:={}\mem&
		y^{\k_1}{\cdots}y^{\k_\ell}\mem
		F^i{}_{j\k_1\s;\k_2;\cdots;\k_\ell}\hnem(x)
    \qiq
    (P_\ell)^i{}_{j\s}\hem y^\s = 0
    \,.
\end{align}
The ancillary file \texttt{P.nb} verifies \eqref{Dpsi-series-expanded} in a direct fashion up to $\O(y^{10})$.
For a physical application,
\eqref{Dpsi} can be used for deriving the Wong equations \cite{wong1970field}
of a color-charged particle
as seen by an arbitrary observer.

To be further concrete, the dressing identity
$\delta^{i\sprime}{}_i\mem \mathe^{\pounds_N}( D\psi^i ) = W^{i\sprime}{}_i\mem (\mathe^{\pounds_N^D} D\psi^i)$ 
unpacks the content of \eqref{Dpsi-series-expanded} as
\begin{align}
	d\psi^{i\sprime}
	+ 
		A^{i\sprime}{}_{j\sprime\r}(z)\mem \psi^{j\sprime}\mem dz^\r
	\,=\,
		W^{i\sprime}{}_i\mem
		\BB{
			d\psi^i 
			+ A^i{}_{j\r}(x)\mem \psi^j\mem dx^\r
		}
		\mem+\,
		W^{i\sprime}{}_i\mem
		\sum_{\ell=1}^\infty\,
			\frac{1}{\ell!}\,
			\BB{
				(P_\ell)^i{}_{j\s}\mem dx^\s
				+ (\ell {\mem-\,} 1)\, (P_{\ell-1})^i{}_{j\s}\mem dy^\s
			}
		\mem \psi^j
	\,,
\end{align}
where the deviated coordinates $z^\m$ is understood as $x^\m + y^\m$.
This reveals that \eqref{Dpsi-series-expanded} encodes
\begin{subequations}
\label{avatar}
\begin{align}
	\label{avatar-a}
	W^i{}_{i'}\hem A^{i\sprime}{}_{j\sprime\mem\r}(z)\mem W^{j\sprime}{}_j
	+
	W^i{}_{i'}\mem \frac{\partial}{\partial x^\r}\mem W^{i\sprime}{}_j
	\,&=\,
		A^i{}_{j\r}(x) + 
		\sum_{\ell=1}^\infty\,
		\frac{1}{\ell!}\,
			(P_\ell)^i{}_{j\r}
	\,,\\
	\label{avatar-b}	
	W^i{}_{i'}\hem A^{i\sprime}{}_{j\sprime\mem\r}(z)\mem W^{j\sprime}{}_j
	+
	W^i{}_{i'}\mem \frac{\partial}{\partial y^\r}\mem W^{i\sprime}{}_j
	\,&=\,
		\sum_{\ell=2}^\infty\,
		\frac{1}{\ell!}\,
			(\ell {\,-\mem} 1)\mem
			(P_{\ell-1})^i{}_{j\r}
	\,.
\end{align}
\end{subequations}
Especially, \eqref{avatar-a} describes
a ``gauge transformation'' of the connection $A^{i\sprime}{}_{j\sprime\mem\r}(z)$
at the deviated point $z$,
via the Wilson line
to the original point $x$.
In this sense, its right-hand side describes an avatar of the gauge connection.
In fact, the sum of $P$-tensors in \eqrefs{avatar-a}{avatar-b},
which computes the difference between 
the connection at $z$ dragged back to $x$
and the connection at $x$,
can be used for studying the Fock-Schwinger gauge \cite{triyanta1991fsgauge}.

Note that all variables in \eqrefs{avatar-a}{avatar-b} are understood as functions of $x$ and $y$, by the very construction of our formalism.
Also, \eqrefs{avatar-a}{avatar-b} can be reproduced
from the following formula
that describes generic variations of the Wilson line
$W(s_2,s_1) = \mathrm{P}\exp\bigbig{ -\hnem\nem\int_{s_1}^{s_2} ds\, A_\r(\gamma(s)\hnem)\mem \dot{\gamma}^\r(s) }$
about an arbitrary contour $s \mapsto \c^\m(s)$:
\begin{align}
\begin{split}
	W(0,1)\mem \delta W(1,0)
	\mem&=\mem
	\begin{aligned}[t]
		&
		A_\s(\gamma(0)\hnem)\mem \delta\gamma^\s(0)
		- W(0,1)\mem 
			A_\s(\gamma(1)\hnem)\mem \delta\gamma^\s(1)
		\hem W(1,0)
		\\
		&
		+ \int_0^1 ds\,\,
			\dot{\gamma}^\r(s)\,
			W(0,s)\mem F_{\r\s}(\gamma(s)\hnem)\mem W(s,0)
			\,\delta\gamma^\s(s)
		\,.
	\end{aligned}
\end{split}
\end{align}

Note that the identities in \eqref{avatar}
are the consequences of the conjugation 
$\mathe^{\pounds_N^D} D\mem \mathe^{-\pounds_N^D}$;
recall the discussion around \eqref{torid}.
In the same fashion,
more identities 
follow by conjugating
$D^2 = F$, $[D , F] = 0$, {etc.}
by \smash{$\mathe^{\pounds_N^D}$}.

Lastly,
the identities analogous to \eqref{avatar}
in Riemannian geometry
are
\begin{subequations}
\label{avatar-gravity}
\begin{align}
	\label{avatar-gravity-a}
		W^\m{}_{\m'}\mem
		\Gamma^{\m\sprime}{}_{\n\sprime\r'}(z)\mem
		W^{\n\sprime}{}_\n
		\, W^{\r\sprime}{}_\k\mem X^\k{}_\r
	+
		W^\m{}_{\m'}\mem
		\bb{
			\frac{\partial}{\partial x^\r}
			- \Gamma^\k{}_{\l\r}(x)\mem 
				y^\l
				\frac{\partial}{\partial y^\k}
		}\mem 
			W^{\m\sprime}{}_\n
	\,&=\,
		\Gamma^\m{}_{\n\r}(x)
		+ \acX^\m{}_{\n\r}
	\,,\\
	\label{avatar-gravity-b}
		W^\m{}_{\m'}\mem
		\Gamma^{\m\sprime}{}_{\n\sprime\r'}(z)\mem
		W^{\n\sprime}{}_\n
		\, W^{\r\sprime}{}_\k\mem Y^\k{}_\r
	+
		W^\m{}_{\m'}\mem
		\frac{\partial}{\partial y^\r}\mem 
			W^{\m\sprime}{}_\n
	\,&=\,
		\acY^\m{}_{\n\r}
	\,,
\end{align}
\end{subequations}
which can be deduced from \eqref{dressing-Dp}.
Notice the presence of the Jacobi propagators $X^\k{}_\r$ and $Y^\k{}_\r$
on the left-hand sides,
which reflects the fact that
the curve for the Wilson lines
depends on the gravitational fields
unlike as in gauge theory.

\twocolumngrid
\pagebreak

\bibliography{references.bib}

\end{document}